\documentclass[reprint,prl,aps,twocolumn,superscriptaddress]{revtex4-1}

\usepackage{bm}
\usepackage{graphicx}
\usepackage{times}
\usepackage{soul}
\usepackage{color}

\usepackage[colorlinks,
            citecolor = blue, 
						linkcolor = blue,
						urlcolor  = blue,
						anchorcolor = blue,
						breaklinks = true
						]{hyperref}

\def \ETH{Institute for Quantum Electronics, ETH Z\"urich, CH-8093 Z\"urich, Switzerland}
\def \UW{Institute of Experimental Physics, Faculty of Physics, University of Warsaw, Pasteura 5, 02-093 Warsaw, Poland}
\def \NIMS{National Institute for Materials Science, Tsukuba, Ibaraki 305-0044, Japan}

\begin{document}

\title{Shubnikov-de Haas oscillations in optical conductivity of monolayer MoSe$_2$}

\author{\surname{T. Smole\'nski}}\affiliation{\ETH}\affiliation{\UW}
\author{\surname{O. Cotlet}}\affiliation{\ETH}
\author{\surname{A. Popert}}\affiliation{\ETH}
\author{\surname{P. Back}}\affiliation{\ETH}
\author{\surname{Y. Shimazaki}}\affiliation{\ETH}
\author{\surname{P.~Kn\"uppel}}\affiliation{\ETH}
\author{\surname{N.~Dietler}}\affiliation{\ETH}
\author{\surname{T. Taniguchi}}\affiliation{\NIMS}
\author{\surname{K. Watanabe}}\affiliation{\NIMS}
\author{\surname{M. Kroner}}\affiliation{\ETH}
\author{\surname{A. Imamoglu}}\email{imamoglu@phys.ethz.ch}\affiliation{\ETH}

\begin{abstract}
We report polarization-resolved resonant reflection spectroscopy of a charge-tunable atomically-thin valley semiconductor hosting tightly bound excitons coupled to a dilute system of fully spin- and valley-polarized holes in the presence of a strong magnetic field. We find that exciton-hole interactions manifest themselves in hole-density dependent, Shubnikov-de Haas-like oscillations in the energy and line broadening of the excitonic resonances. These oscillations are evidenced to be precisely correlated with the occupation of Landau levels, thus demonstrating that strong interactions between the excitons and Landau-quantized itinerant carriers enable optical investigation of quantum-Hall physics in transition metal dichalcogenides.
\end{abstract}

\maketitle 

Optical excitations of a semiconductor hosting a two-dimensional electron (2DES) or hole (2DHS) system subjected to a strong magnetic field provide a direct tool for investigating quantum-Hall states arising from Landau-level (LL) quantization of the carrier orbital motion~\cite{Nurmikko_PhysToday_1993,Kukushkin_AdvPhys_1996,Potemski_PhysB_1998}. Over the last three decades, this approach has been widely applied to explore a plethora of fascinating many-body phenomena in III-V or II-VI quantum-well heterostructures, such as, modulation of the screening responsible for magneto-oscillations of the excitonic luminescence at integer~\cite{Goldberg_PRL_1990,Gravier_PRL_1998} and fractional filling factors~$\nu$~\cite{Heiman_PRL_1988,Turberfield_PRL_1990,Buhmann_PRL_1990}, or formation of collective spin excitations (Skyrmions) around $\nu=1$~\cite{Aifer_PRL_1996,Plochocka_PRL_2009}.

Qualitatively new avenues for the magneto-optical studies of 2DES or 2DHS have emerged with the advent of atomically-thin transition metal dichalcogenides (TMD), which are direct band-gap semiconductors with two non-equivalent valleys of conduction and valence bands appearing at the $K^\pm$ points of the hexagonal Brillouin zone~\cite{Mak_PRL_2010,Xu_NatPhys_2014,Manzeli_NatRevMater_2017,Mak_NatPhoton_2018}. Owing to heavy effective carrier masses and reduced screening, the cyclotron energy in a TMD monolayer is much smaller than the binding energies of the exciton (0.5~eV) or trion (30~meV)~\cite{Chernikov_PRL_2014,Robert_PRM_2018,Stier_PRL_2018} even at high magnetic fields $\sim$10~T. In this regard, TMD monolayers remain in stark contrast to conventional semiconductors, such as GaAs. Moreover, spin-valley locking by the spin-orbit coupling~\cite{Xiao_PRL_2012} as well as the presence of the valley-contrasting $\pi$ Berry phase~\cite{Xu_NatPhys_2014} give rise to a unique ladder of spin- and valley-polarized LLs in TMD monolayers~\cite{Cai_PRB_2013,Rose_PRB_2013}, the degeneracy of which is further lifted by the strong Zeeman effect~\cite{Srivastava_NP_2015,Aivazian_NP_2015}. Although the fingerprints of such LLs have been recently reported in several transport experiments~\cite{Fallahazad_PRL_2016,Movva_PRL_2017,Gustafsson_NatMater_2018,Larentis_PRB_2018,Lin_arXiv_2018,Pisoni_arXiv_2018}, their optical signatures have been obtained only for the WSe$_2$ monolayer at high electron densities, where exciton binding is suppressed due to screening; in this limit, the optical excitation spectrum is similar to that of GaAs 2DES at moderate fields and shows band-to-band inter-LL transitions~\cite{Wang_NatNano_2017}. 

In this Letter, we report the optical response of a Landau-quantized 2DHS in a MoSe$_2$ monolayer placed in a strong magnetic field of 8--16~T in the opposite limit of \emph{low} densities \mbox{$p\lesssim3\cdot10^{12}$~cm$^{-2}$}, where interaction effects are manifest. For such low hole densities (corresponding to a Wigner-Seitz radius of $r_s\gtrsim5$), the excitons remain tightly bound and the itinerant holes are fully spin- and valley-polarized in the $K^+$ valley (for $B>0$). As a consequence, the optical excitation spectrum is dominated by two cross-circularly-polarized resonances corresponding to the bare exciton in the $K^+$ valley and exciton-polaron in the $K^-$ valley~\cite{Sidler_NatPhys_2017,Efimkin_PRB_2017,Back_PRL_2018,Efimkin_PRB_2018}. We demonstrate that the transition energies and/or linewidths of both resonances exhibit Shubnikov-de Haas-like (SdH-like) oscillations with the hole density, which are correlated with LL filling, thus providing a first direct evidence for the influence of quantum-Hall states on the excitonic excitations in a TMD monolayer.

\begin{figure}
\includegraphics{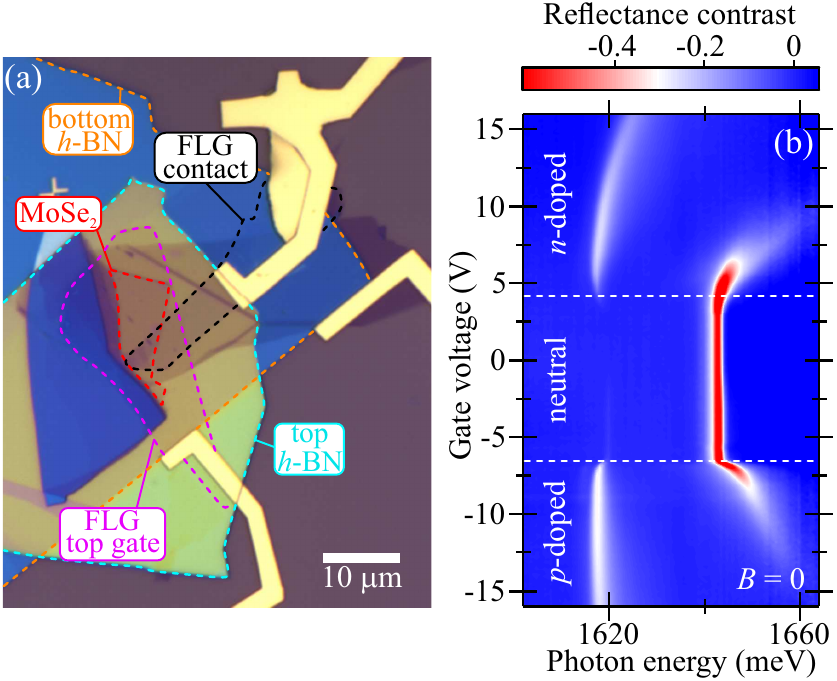}
\caption{(a)~Optical microscope image of the gate-controlled MoSe$_2$ monolayer studied in this work. The monolayer was encapsulated between two layers of $h$-BN, electrically contacted with a few-layer graphene (FLG) flake, and capped with the second FLG flake serving as a top gate (the boundaries of the flakes are marked with dashed lines). The carrier density was tuned by applying a gate voltage between Au/Ti electrodes connected to the FLG flakes. (b)~Color-scale map presenting reflectance contrast spectra measured as a function of the gate voltage at $B=0$. The horizontal dashed lines mark the transitions between neutral and $n$- or $p$-doped regimes.\label{fig:sample_and_0T_refl_different_map}}
\end{figure}

Our experiments have been carried out on three different van der Waals heterostructures, each consisting of a~gate-controlled MoSe$_2$ monolayer encapsulated between two hexagonal boron-nitride ($h$-BN) layers. In the main text, we present the data acquired for one of these devices [Fig.~\ref{fig:sample_and_0T_refl_different_map}(a)], where density-dependent oscillations are manifest in the hole-doped regime; while all 3 devices show such oscillations in the presence of a 2DHS, the third device, described in the Supplemental Material (SM)~\cite{SM}, displays analogous oscillations for a 2DES. The valley-selective optical response of all samples has been analyzed by means of low-temperature ($T\approx4$~K), circular-polarization-resolved, white light reflection magneto-spectroscopy (for details, see SM~\cite{SM}).

We first concentrate on the spectral reflectance contrast measured as a function of the gate voltage $V_g$ at $B=0$, which is shown in Fig~\ref{fig:sample_and_0T_refl_different_map}(b). For $-7\ \mathrm{V}\lesssim V_g\lesssim4\ \mathrm{V}$, when the sample is devoid of free carriers, the spectrum features only the bare exciton resonance. Once $V_g$ is increased (decreased) beyond this range, free electrons (holes) start to be injected to the monolayer. As demonstrated previously~\cite{Sidler_NatPhys_2017,Efimkin_PRB_2017,Back_PRL_2018,Efimkin_PRB_2018}, the attractive interaction between these carriers and the excitons dress the latter into exciton-polarons, which in turn qualitatively alters the nature of the optical transitions. A~prominent signature of this crossover is an emergence of a second, lower-energy resonance originating from an attractive exciton-polaron. Concurrently, the exciton resonance transforms into a repulsive exciton-polaron resonance, which exhibits a strong blue-shift and line broadening. Due to the transfer of oscillator strength to the attractive polaron, the repulsive polaron resonance becomes indiscernible for carrier densities~$\gtrsim1\cdot10^{12}$~cm$^{-2}$.

\begin{figure}[b]
\includegraphics{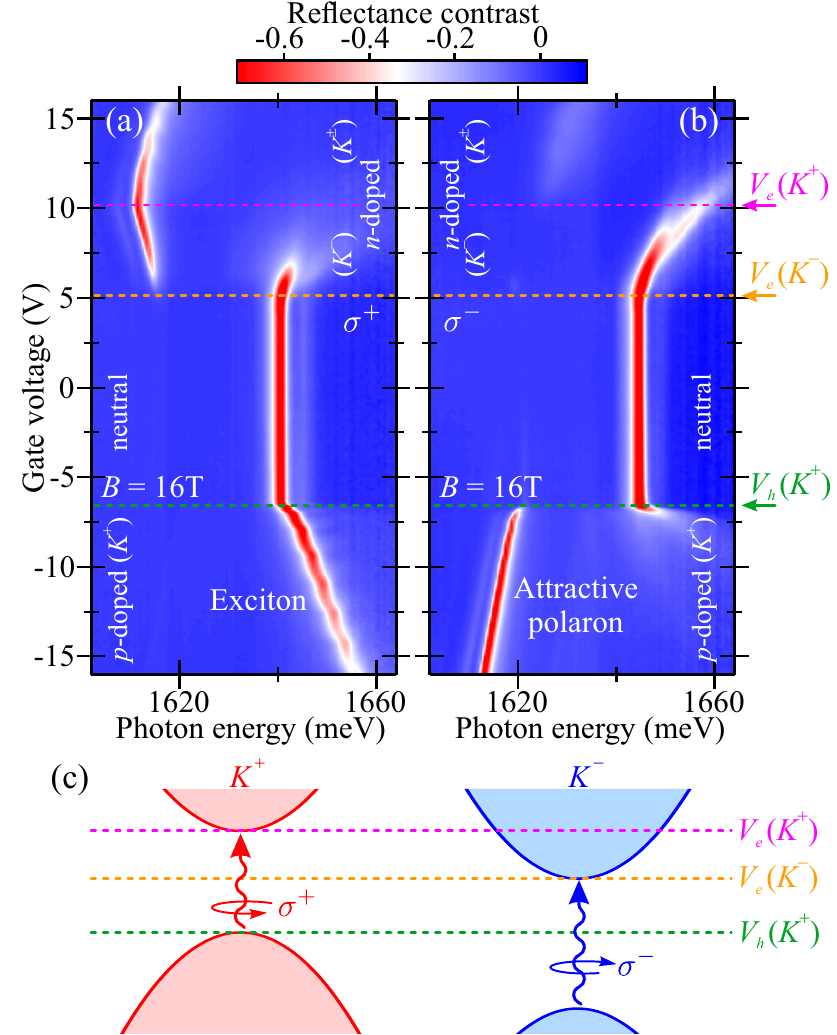}
\caption{(a,b) Reflectance contrast spectra measured at $B=16$~T as a~function of the gate voltage in $\sigma^+$ (a) or $\sigma^-$ (b) circular polarization. (c) Schematic illustrating the lowest-energy electronic subbands at the $K^+$ and $K^-$ valleys for a MoSe$_2$ monolayer subjected to an external magnetic field. The horizontal dashed lines show the positions of the Fermi level at the crossovers between different doping regimes, which are marked in (a,b).\label{fig:16T_refl_and_cartoon}}
\end{figure}

\begin{figure*}
\includegraphics{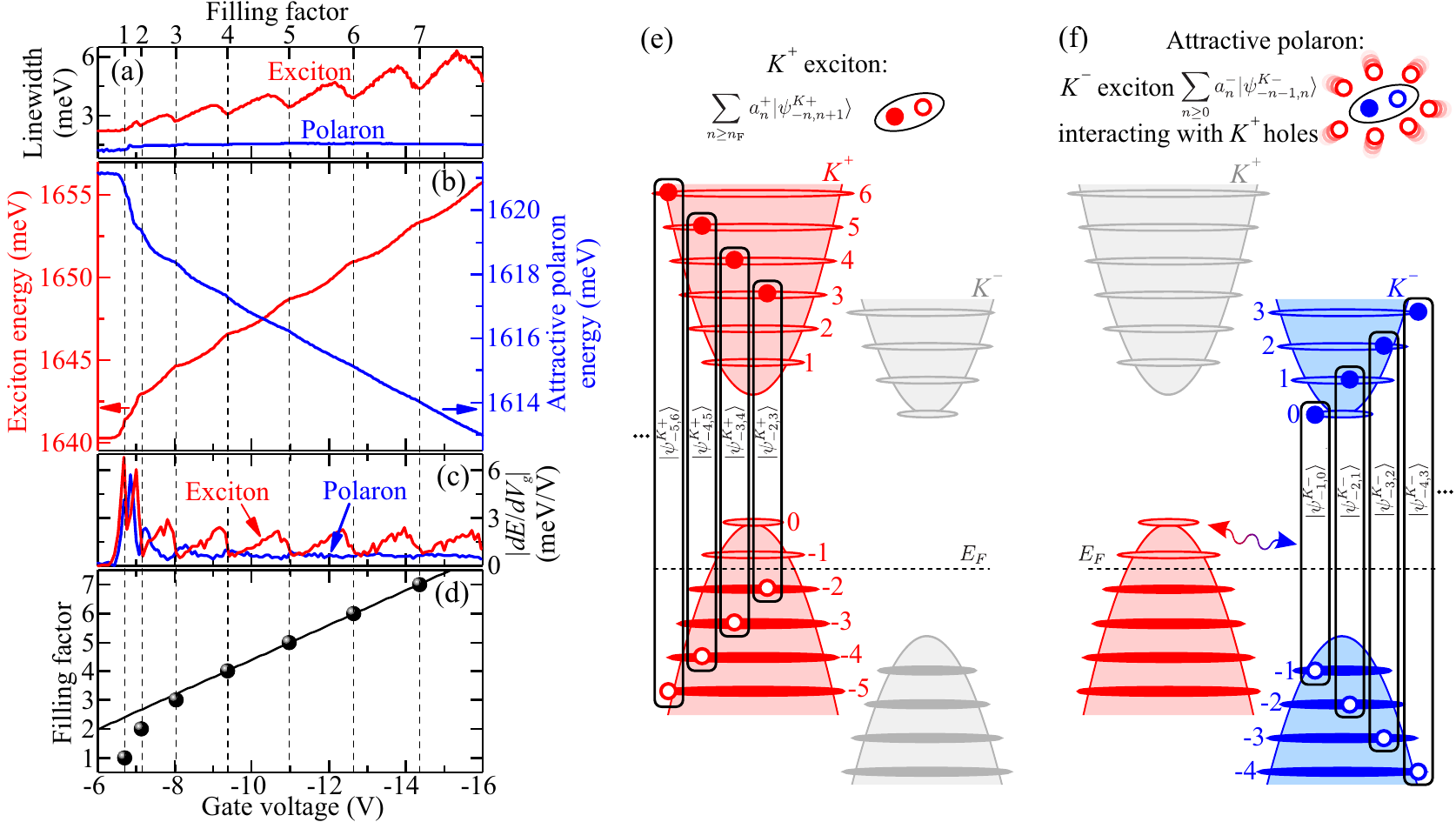}
\caption{(a,b)~Gate-voltage dependencies of the linewidths~(a) and energies~(b) of the exciton (red) and attractive polaron (blue) resonances on the hole-doped side at $B=16$~T. The data was extracted by fitting the spectral profiles of both resonances in the polarization-resolved reflectance contrast spectra from Figs~\ref{fig:16T_refl_and_cartoon}(a,b). (c)~Absolute values of the derivatives of the energies of both resonances with respect to the gate voltage. (d)~Gate voltages $V_g$ corresponding to integer filling factors $\nu$ determined based on the positions of the local minima of the exciton linewidth (independently marked by vertical dashed lines). Solid line represents the linear fit to the data for $V_g\lesssim-8$~V, when the Schottky effects at the contact to MoSe$_2$ can be neglected, and hence when the hole density depends linearly on $V_g$ (for details, see SM~\cite{SM}). (e,f)~Cartoons illustrating the exciton~(e) and attractive polaron~(f) configurations in the presence of Landau-quantized system of spin-polarized holes occupying only the states in $K^+$ valley under the influence of external magnetic field $B>0$.\label{fig:exciton_apolaron_LLs}}
\end{figure*}

Figs~\ref{fig:16T_refl_and_cartoon}(a,b) present the gate-voltage-dependence of the reflectance contrast spectra detected in two circular polarizations upon application of the magnetic field of $B=16$~T perpendicularly to the monolayer plane. Such a field lifts the degeneracy of the electronic states in $K^\pm$ valleys due to the Zeeman splitting [see Fig.~\ref{fig:16T_refl_and_cartoon}(c)], whose magnitude has been previously found to be significantly enhanced by the exchange interactions in the presence of free carriers~\cite{Back_PRL_2017}. Consequently, at appropriately low densities the electrons (holes) occupy only the states in $K^-$ ($K^+$) valley. Because the polaron dressing of the excitons arises predominantly due to the intervalley exciton-carrier interaction~\cite{Efimkin_PRB_2017,Back_PRL_2017,Efimkin_PRB_2018}, at such densities the exciton-polaron resonances appear exclusively in $\sigma^+$ ($\sigma^-$) polarization, whereas the spectrum in the opposite polarization shows only the bare exciton resonance. As seen in Figs~\ref{fig:16T_refl_and_cartoon}(a,b), this valley-polarized regime covers the whole experimentally accessible gate-voltage range $-16\ \mathrm{V}\lesssim V_g\lesssim\ -7\ \mathrm{V}$ on the hole-doped side, while for the electrons it holds for $5\ \mathrm{V}\lesssim V_g\lesssim\ 10\ \mathrm{V}$. Most importantly, there is a striking difference between these two cases: at electron doping the optical transitions evolve smoothly with $V_g$, whereas at hole doping they exhibit a pronounced oscillatory behavior. As we show below, these oscillations are due to sequential filling of the hole LLs. 

Although we have also observed similar oscillations on the electron-doped side for a better quality sample featuring narrower optical transitions (see SM~\cite{SM}), they have been found to be much less pronounced than the oscillations at the hole doping for all studied devices. We speculate that this asymmetry may be, at least partially, a consequence of electron-to-hole effective mass ratio that exceeds unity, in contrast to density functional theory (DFT) calculations predicting both masses to be similar~\cite{Cheiwchanchamnangij_PRB_2012,Yun_PRB_2012,Kormanyos_2DMat_2015}. This conjecture is supported by the fact that while the hole mass was shown by ARPES measurements~\cite{Jin_PRB_2015,Tanabe_APL_2016,Ulstrup_ACSNano_2016,Wilson_SciAdv_2017} to remain in excellent agreement with DFT, recent transport measurements of MoS$_2$ and MoSe$_2$ monolayers indicated that the electron mass may be larger by a factor of~$\sim2$~\cite{Larentis_PRB_2018,Pisoni_arXiv_2018}.

In the following we focus on the central finding of our work---the signatures of LL filling in the optical spectra of the exciton and attractive polaron transitions in the hole-doped regime (the repulsive polaron transition is not examined, since it disappears at very low hole densities). As seen in \mbox{Figs~\ref{fig:exciton_apolaron_LLs}(a-c)}, both resonances exhibit aforementioned oscillatory behavior, which is particularly striking for the exciton, whose linewidth displays sharp, periodic minima [Fig.~\ref{fig:exciton_apolaron_LLs}(a)] that appear to be correlated with cusp-like changes of the slope of the energy increase with~$V_g$ [Figs~\ref{fig:exciton_apolaron_LLs}(b,c)]. Remarkably, this effect may be independently understood as SdH oscillations in MoSe$_2$ optical conductivity~$\sigma(E_\mathrm{X})$ at the excitonic energy~\cite{Chen_PRL_1990}, since $\mathrm{Re}[\sigma(E_\mathrm{X})]$ is determined by the excitonic linewidth~\cite{Scuri_PRL_2018}. However, the origin of those oscillations is different to that of transport SdH oscillations in static conductivity. Specifically, we find that the oscillations in our system are due to the influence of the LL occupation on the strength of the interactions between the exciton and Landau-quantized 2DHS. This is particularly clear for the case of the exciton linewidth, the narrowing of which arises directly from the modulation of the efficiency of the intravalley exciton-hole coupling. This coupling becomes enhanced each time the lowest-energy LL is only partially filled (around half-integer~$\nu$), as in such a case the holes can be effectively scattered between the empty states belonging to this LL. Concurrently, when the Fermi level lies in the gap between the LLs (around integer~$\nu$), the possibility of such a hole scattering is substantially reduced due to lack of available final states. This leads to a suppression of the exciton-hole interaction, and thereby gives rise to a pronounced minima in the exciton linewidth.

\begin{figure}
\includegraphics{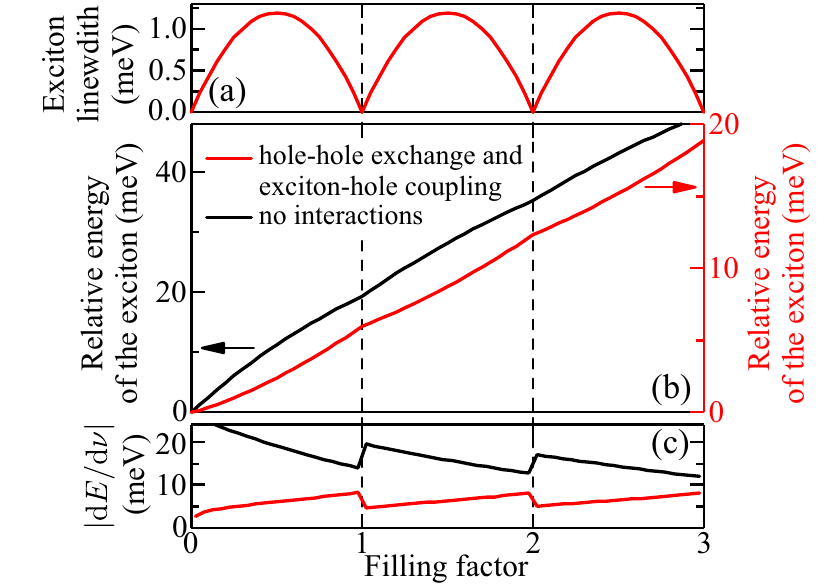}
\caption{(a-c)~Exciton linewidth~(a), relative energy~(b), and derivative of the energy with respect to~$\nu$~(c) calculated at $B=16$~T as a~function of the filling factor using the developed theoretical model. The red (black) curves represent the results of the model including (excluding) the interaction effects.\label{fig:theory}}
\end{figure}

The above reasoning is corroborated by the results of our theoretical simulations of the exciton absorption spectrum, which, for simplicity, were intended to reproduce the LL-related oscillatory features on a qualitative level only. First, using a variational approach, we calculate the energy of the exciton embedded in the reservoir of valley-polarized holes, taking into account both the phase-space filling and hole-hole exchange interaction. Then, assuming contact-like repulsive intravalley exciton-hole interaction, we calculate (to second order in the interaction) the correlation energy and finite lifetime acquired by the exciton due to dressing with electron-hole pairs from the Fermi sea (for details, see SM~\cite{SM}). As shown in Figs~\ref{fig:theory}(a-c), this simple model correctly captures periodic narrowing of the exciton transition around integer~$\nu$. Moreover, it also predicts the slope of exciton energy dependence to be altered for the same hole densities, thus explaining the presence of cusp-like features in Figs~\ref{fig:exciton_apolaron_LLs}(b,c). We emphasize that although analogous oscillatory features appear in the model including only the phase-space filling and ignoring interactions [see the black curves in~Figs~\ref{fig:theory}(b,c)], in such a case the cusps have an opposite direction to that in the experiment: the slope of the exciton energy gets steeper around integer $\nu$ instead of becoming flatter. This finding unequivocally demonstrates that the correct description of the experimental data requires inclusion of the interactions, which in turn confirms their primary role in LL-filling-dependent modification of excitonic spectra.

Remarkably, the signatures of LL-filling emerge also for the $K^-$ exciton dressed into an attractive polaron, but are found to be much less pronounced. In fact, they are barely visible in the gate-voltage evolution of the polaron linewidth [Fig.~\ref{fig:exciton_apolaron_LLs}(a)] and appear only in the $V_g$-dependence of the energy, taking form of a familiar cusp-like features around integer $\nu$ that, however, tend to vanish for larger hole densities [Fig.~\ref{fig:exciton_apolaron_LLs}(b,c)]. Although the reason behind this tendency remains not entirely clear, the presence of even faint polaron energy oscillations is by itself supportive of our explanation of the LL signatures to emerge in the optical spectra due to the interactions. This stems from the fact that, unlike the exciton, the polaron transition is not affected by the phase-space filling in the investigated valley-polarized regime [compare the schematics in Figs~\ref{fig:exciton_apolaron_LLs}(e,f)].

\begin{figure}
\includegraphics{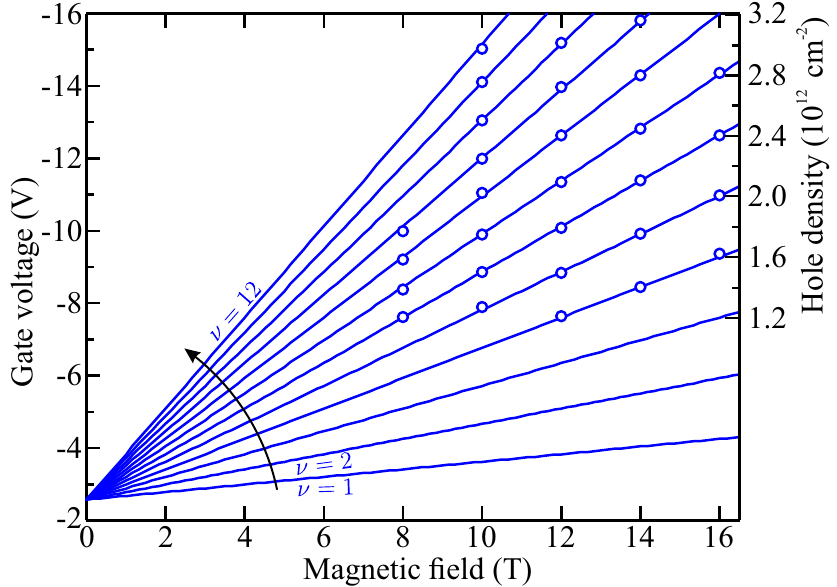}
\caption{Gate voltages $V_g(\nu,B)$ corresponding to integer filling factors $\nu$ determined based on the positions of the exciton linewidth minima for different magnetic fields $B$. The data are presented only in a linear-response regime of the gate (i.e., for $V_g\lesssim-8$~V), in which the hole density depends linearly on $V_g$, and its value (marked on the right axis) can be obtained within the frame of the parallel-plate capacitor model (for details, see SM~\cite{SM}). Solid lines represent the fit to the experimental data with a set of linear dependencies $V_g(\nu,B)=V_0-\delta\nu B$ with two fitting parameters: $V_0$ representing the common origin of these curves, and $\delta$ controlling their slopes.\label{fig:filling_LLs}}
\end{figure}

As demonstrated above, the minima of the exciton linewidth occur at gate voltages corresponding to subsequent integer $\nu$, which should imply these voltages $V_g(\nu,B)$ to be equally spaced, as long as the hole density $p$ remains proportional to $V_g$. Although this prediction holds for $\nu>3$ [see Fig.~\ref{fig:exciton_apolaron_LLs}(d)], at lower densities the inter-LL voltage gaps turn out to be significantly narrower. We attribute these deviations (that are found to be specific to the investigated device, see SM~\cite{SM}) to the presence of a Schottky barrier at the contact to the MoSe$_2$, which gives rise to initially non-linear increase of $p$ with $V_g$ (for details, see SM~\cite{SM}). This effect, however, becomes negligible at larger-density regime (i.e., for $V_g\lesssim-8$~V), where the voltages $V_g(\nu, B)$ extracted at different fields $B$ indeed scale proportionally to both $\nu$ and $B$, as evidenced by their perfect agreement with a fitted set of linear dependencies forming a characteristic LL fan chart (Fig.~\ref{fig:filling_LLs}). The fit allows us to determine the voltage change $\delta\approx0.1$~V/T needed to fill a LL in a unit magnetic field. On this basis, as well as based on the geometrical capacitance $C_\mathrm{geom}$ of our device (obtained within a parallel-plate capacitor approximation, see SM~\cite{SM}), we finally extract the number of states in each LL $C_\mathrm{geom}\delta B/e$ that yields $(1.0\pm0.2)eB/h$ (where $e$ is the electron charge, while $h$ the Planck constant). This finding indicates complete lifting of the LL degeneracy, exactly as expected for LLs that are valley- and spin-polarized.

In conclusion, our experiments provide the first signatures of LL-quantization in the optical spectra of a TMD monolayer hosting a dilute system of spin-valley polarized holes. These signatures are evidenced to emerge due to the influence of LL occupation on the strength of interactions between itinerant holes and tightly bound excitons, which gives rise to prominent filling-factor-dependent SdH-like oscillations in the energy and linewidth of the excitonic transitions. The interaction-enabled optical access to the quantum-Hall physics demonstrated in our work constitutes the first step towards optical investigation of a rich field of strongly correlated phenomena at integer and fractional filling factors in atomically thin semiconductors.

\begin{acknowledgments}
The authors acknowledge discussions with Richard Schmidt. This work is supported by a~European Research Council (ERC) Advanced investigator grant (POLTDES), and by a~grant from Swiss National Science Foundation (SNSF). T.S.~is supported by the Polish National Science Centre through PhD scholarship Grant No. DEC-2016/20/T/ST3/00028. K.W. and T.T. acknowledge support from the Elemental Strategy Initiative conducted by the MEXT, Japan and the CREST (JPMJCR15F3), JST.
\end{acknowledgments}


\begin{thebibliography}{44}%
\makeatletter
\providecommand \@ifxundefined [1]{%
 \@ifx{#1\undefined}
}%
\providecommand \@ifnum [1]{%
 \ifnum #1\expandafter \@firstoftwo
 \else \expandafter \@secondoftwo
 \fi
}%
\providecommand \@ifx [1]{%
 \ifx #1\expandafter \@firstoftwo
 \else \expandafter \@secondoftwo
 \fi
}%
\providecommand \natexlab [1]{#1}%
\providecommand \enquote  [1]{``#1''}%
\providecommand \bibnamefont  [1]{#1}%
\providecommand \bibfnamefont [1]{#1}%
\providecommand \citenamefont [1]{#1}%
\providecommand \href@noop [0]{\@secondoftwo}%
\providecommand \href [0]{\begingroup \@sanitize@url \@href}%
\providecommand \@href[1]{\@@startlink{#1}\@@href}%
\providecommand \@@href[1]{\endgroup#1\@@endlink}%
\providecommand \@sanitize@url [0]{\catcode `\\12\catcode `\$12\catcode
  `\&12\catcode `\#12\catcode `\^12\catcode `\_12\catcode `\%12\relax}%
\providecommand \@@startlink[1]{}%
\providecommand \@@endlink[0]{}%
\providecommand \url  [0]{\begingroup\@sanitize@url \@url }%
\providecommand \@url [1]{\endgroup\@href {#1}{\urlprefix }}%
\providecommand \urlprefix  [0]{URL }%
\providecommand \Eprint [0]{\href }%
\providecommand \doibase [0]{http://dx.doi.org/}%
\providecommand \selectlanguage [0]{\@gobble}%
\providecommand \bibinfo  [0]{\@secondoftwo}%
\providecommand \bibfield  [0]{\@secondoftwo}%
\providecommand \translation [1]{[#1]}%
\providecommand \BibitemOpen [0]{}%
\providecommand \bibitemStop [0]{}%
\providecommand \bibitemNoStop [0]{.\EOS\space}%
\providecommand \EOS [0]{\spacefactor3000\relax}%
\providecommand \BibitemShut  [1]{\csname bibitem#1\endcsname}%
\let\auto@bib@innerbib\@empty
%</preamble>
\bibitem [{\citenamefont {Nurmikko}\ and\ \citenamefont
  {Pinczuk}(1993)}]{Nurmikko_PhysToday_1993}%
  \BibitemOpen
  \bibfield  {author} {\bibinfo {author} {\bibfnamefont {A.}~\bibnamefont
  {Nurmikko}}\ and\ \bibinfo {author} {\bibfnamefont {A.}~\bibnamefont
  {Pinczuk}},\ }\href {\doibase 10.1063/1.881352} {\bibfield  {journal}
  {\bibinfo  {journal} {Phys. Today}\ }\textbf {\bibinfo {volume} {46}},\
  \bibinfo {pages} {24} (\bibinfo {year} {1993})}\BibitemShut {NoStop}%
\bibitem [{\citenamefont {Kukushkin}\ and\ \citenamefont
  {Timofeev}(1996)}]{Kukushkin_AdvPhys_1996}%
  \BibitemOpen
  \bibfield  {author} {\bibinfo {author} {\bibfnamefont {I.~V.}\ \bibnamefont
  {Kukushkin}}\ and\ \bibinfo {author} {\bibfnamefont {V.~B.}\ \bibnamefont
  {Timofeev}},\ }\href {\doibase 10.1080/00018739600101487} {\bibfield
  {journal} {\bibinfo  {journal} {Adv. Phys.}\ }\textbf {\bibinfo {volume}
  {45}},\ \bibinfo {pages} {147} (\bibinfo {year} {1996})}\BibitemShut
  {NoStop}%
\bibitem [{\citenamefont {Potemski}(1998)}]{Potemski_PhysB_1998}%
  \BibitemOpen
  \bibfield  {author} {\bibinfo {author} {\bibfnamefont {M.}~\bibnamefont
  {Potemski}},\ }\href {\doibase 10.1016/S0921-4526(98)00566-3} {\bibfield
  {journal} {\bibinfo  {journal} {Phys. B}\ }\textbf {\bibinfo {volume}
  {256--258}},\ \bibinfo {pages} {283} (\bibinfo {year} {1998})}\BibitemShut
  {NoStop}%
\bibitem [{\citenamefont {Goldberg}\ \emph {et~al.}(1990)\citenamefont
  {Goldberg}, \citenamefont {Heiman}, \citenamefont {Pinczuk}, \citenamefont
  {Pfeiffer},\ and\ \citenamefont {West}}]{Goldberg_PRL_1990}%
  \BibitemOpen
  \bibfield  {author} {\bibinfo {author} {\bibfnamefont {B.~B.}\ \bibnamefont
  {Goldberg}}, \bibinfo {author} {\bibfnamefont {D.}~\bibnamefont {Heiman}},
  \bibinfo {author} {\bibfnamefont {A.}~\bibnamefont {Pinczuk}}, \bibinfo
  {author} {\bibfnamefont {L.}~\bibnamefont {Pfeiffer}}, \ and\ \bibinfo
  {author} {\bibfnamefont {K.}~\bibnamefont {West}},\ }\href {\doibase
  10.1103/PhysRevLett.65.641} {\bibfield  {journal} {\bibinfo  {journal} {Phys.
  Rev. Lett.}\ }\textbf {\bibinfo {volume} {65}},\ \bibinfo {pages} {641}
  (\bibinfo {year} {1990})}\BibitemShut {NoStop}%
\bibitem [{\citenamefont {Gravier}\ \emph {et~al.}(1998)\citenamefont
  {Gravier}, \citenamefont {Potemski}, \citenamefont {Hawrylak},\ and\
  \citenamefont {Etienne}}]{Gravier_PRL_1998}%
  \BibitemOpen
  \bibfield  {author} {\bibinfo {author} {\bibfnamefont {L.}~\bibnamefont
  {Gravier}}, \bibinfo {author} {\bibfnamefont {M.}~\bibnamefont {Potemski}},
  \bibinfo {author} {\bibfnamefont {P.}~\bibnamefont {Hawrylak}}, \ and\
  \bibinfo {author} {\bibfnamefont {B.}~\bibnamefont {Etienne}},\ }\href
  {\doibase 10.1103/PhysRevLett.80.3344} {\bibfield  {journal} {\bibinfo
  {journal} {Phys. Rev. Lett.}\ }\textbf {\bibinfo {volume} {80}},\ \bibinfo
  {pages} {3344} (\bibinfo {year} {1998})}\BibitemShut {NoStop}%
\bibitem [{\citenamefont {Heiman}\ \emph {et~al.}(1988)\citenamefont {Heiman},
  \citenamefont {Goldberg}, \citenamefont {Pinczuk}, \citenamefont {Tu},
  \citenamefont {Gossard},\ and\ \citenamefont {English}}]{Heiman_PRL_1988}%
  \BibitemOpen
  \bibfield  {author} {\bibinfo {author} {\bibfnamefont {D.}~\bibnamefont
  {Heiman}}, \bibinfo {author} {\bibfnamefont {B.~B.}\ \bibnamefont
  {Goldberg}}, \bibinfo {author} {\bibfnamefont {A.}~\bibnamefont {Pinczuk}},
  \bibinfo {author} {\bibfnamefont {C.~W.}\ \bibnamefont {Tu}}, \bibinfo
  {author} {\bibfnamefont {A.~C.}\ \bibnamefont {Gossard}}, \ and\ \bibinfo
  {author} {\bibfnamefont {J.~H.}\ \bibnamefont {English}},\ }\href {\doibase
  10.1103/PhysRevLett.61.605} {\bibfield  {journal} {\bibinfo  {journal} {Phys.
  Rev. Lett.}\ }\textbf {\bibinfo {volume} {61}},\ \bibinfo {pages} {605}
  (\bibinfo {year} {1988})}\BibitemShut {NoStop}%
\bibitem [{\citenamefont {Turberfield}\ \emph {et~al.}(1990)\citenamefont
  {Turberfield}, \citenamefont {Haynes}, \citenamefont {Wright}, \citenamefont
  {Ford}, \citenamefont {Clark}, \citenamefont {Ryan}, \citenamefont {Harris},\
  and\ \citenamefont {Foxon}}]{Turberfield_PRL_1990}%
  \BibitemOpen
  \bibfield  {author} {\bibinfo {author} {\bibfnamefont {A.~J.}\ \bibnamefont
  {Turberfield}}, \bibinfo {author} {\bibfnamefont {S.~R.}\ \bibnamefont
  {Haynes}}, \bibinfo {author} {\bibfnamefont {P.~A.}\ \bibnamefont {Wright}},
  \bibinfo {author} {\bibfnamefont {R.~A.}\ \bibnamefont {Ford}}, \bibinfo
  {author} {\bibfnamefont {R.~G.}\ \bibnamefont {Clark}}, \bibinfo {author}
  {\bibfnamefont {J.~F.}\ \bibnamefont {Ryan}}, \bibinfo {author}
  {\bibfnamefont {J.~J.}\ \bibnamefont {Harris}}, \ and\ \bibinfo {author}
  {\bibfnamefont {C.~T.}\ \bibnamefont {Foxon}},\ }\href {\doibase
  10.1103/PhysRevLett.65.637} {\bibfield  {journal} {\bibinfo  {journal} {Phys.
  Rev. Lett.}\ }\textbf {\bibinfo {volume} {65}},\ \bibinfo {pages} {637}
  (\bibinfo {year} {1990})}\BibitemShut {NoStop}%
\bibitem [{\citenamefont {Buhmann}\ \emph {et~al.}(1990)\citenamefont
  {Buhmann}, \citenamefont {Joss}, \citenamefont {von Klitzing}, \citenamefont
  {Kukushkin}, \citenamefont {Martinez}, \citenamefont {Plaut}, \citenamefont
  {Ploog},\ and\ \citenamefont {Timofeev}}]{Buhmann_PRL_1990}%
  \BibitemOpen
  \bibfield  {author} {\bibinfo {author} {\bibfnamefont {H.}~\bibnamefont
  {Buhmann}}, \bibinfo {author} {\bibfnamefont {W.}~\bibnamefont {Joss}},
  \bibinfo {author} {\bibfnamefont {K.}~\bibnamefont {von Klitzing}}, \bibinfo
  {author} {\bibfnamefont {I.~V.}\ \bibnamefont {Kukushkin}}, \bibinfo {author}
  {\bibfnamefont {G.}~\bibnamefont {Martinez}}, \bibinfo {author}
  {\bibfnamefont {A.~S.}\ \bibnamefont {Plaut}}, \bibinfo {author}
  {\bibfnamefont {K.}~\bibnamefont {Ploog}}, \ and\ \bibinfo {author}
  {\bibfnamefont {V.~B.}\ \bibnamefont {Timofeev}},\ }\href {\doibase
  10.1103/PhysRevLett.65.1056} {\bibfield  {journal} {\bibinfo  {journal}
  {Phys. Rev. Lett.}\ }\textbf {\bibinfo {volume} {65}},\ \bibinfo {pages}
  {1056} (\bibinfo {year} {1990})}\BibitemShut {NoStop}%
\bibitem [{\citenamefont {Aifer}\ \emph {et~al.}(1996)\citenamefont {Aifer},
  \citenamefont {Goldberg},\ and\ \citenamefont {Broido}}]{Aifer_PRL_1996}%
  \BibitemOpen
  \bibfield  {author} {\bibinfo {author} {\bibfnamefont {E.~H.}\ \bibnamefont
  {Aifer}}, \bibinfo {author} {\bibfnamefont {B.~B.}\ \bibnamefont {Goldberg}},
  \ and\ \bibinfo {author} {\bibfnamefont {D.~A.}\ \bibnamefont {Broido}},\
  }\href {\doibase 10.1103/PhysRevLett.76.680} {\bibfield  {journal} {\bibinfo
  {journal} {Phys. Rev. Lett.}\ }\textbf {\bibinfo {volume} {76}},\ \bibinfo
  {pages} {680} (\bibinfo {year} {1996})}\BibitemShut {NoStop}%
\bibitem [{\citenamefont {Plochocka}\ \emph {et~al.}(2009)\citenamefont
  {Plochocka}, \citenamefont {Schneider}, \citenamefont {Maude}, \citenamefont
  {Potemski}, \citenamefont {Rappaport}, \citenamefont {Umansky}, \citenamefont
  {Bar-Joseph}, \citenamefont {Groshaus}, \citenamefont {Gallais},\ and\
  \citenamefont {Pinczuk}}]{Plochocka_PRL_2009}%
  \BibitemOpen
  \bibfield  {author} {\bibinfo {author} {\bibfnamefont {P.}~\bibnamefont
  {Plochocka}}, \bibinfo {author} {\bibfnamefont {J.~M.}\ \bibnamefont
  {Schneider}}, \bibinfo {author} {\bibfnamefont {D.~K.}\ \bibnamefont
  {Maude}}, \bibinfo {author} {\bibfnamefont {M.}~\bibnamefont {Potemski}},
  \bibinfo {author} {\bibfnamefont {M.}~\bibnamefont {Rappaport}}, \bibinfo
  {author} {\bibfnamefont {V.}~\bibnamefont {Umansky}}, \bibinfo {author}
  {\bibfnamefont {I.}~\bibnamefont {Bar-Joseph}}, \bibinfo {author}
  {\bibfnamefont {J.~G.}\ \bibnamefont {Groshaus}}, \bibinfo {author}
  {\bibfnamefont {Y.}~\bibnamefont {Gallais}}, \ and\ \bibinfo {author}
  {\bibfnamefont {A.}~\bibnamefont {Pinczuk}},\ }\href {\doibase
  10.1103/PhysRevLett.102.126806} {\bibfield  {journal} {\bibinfo  {journal}
  {Phys. Rev. Lett.}\ }\textbf {\bibinfo {volume} {102}},\ \bibinfo {pages}
  {126806} (\bibinfo {year} {2009})}\BibitemShut {NoStop}%
\bibitem [{\citenamefont {Mak}\ \emph {et~al.}(2010)\citenamefont {Mak},
  \citenamefont {Lee}, \citenamefont {Hone}, \citenamefont {Shan},\ and\
  \citenamefont {Heinz}}]{Mak_PRL_2010}%
  \BibitemOpen
  \bibfield  {author} {\bibinfo {author} {\bibfnamefont {K.~F.}\ \bibnamefont
  {Mak}}, \bibinfo {author} {\bibfnamefont {C.}~\bibnamefont {Lee}}, \bibinfo
  {author} {\bibfnamefont {J.}~\bibnamefont {Hone}}, \bibinfo {author}
  {\bibfnamefont {J.}~\bibnamefont {Shan}}, \ and\ \bibinfo {author}
  {\bibfnamefont {T.~F.}\ \bibnamefont {Heinz}},\ }\href {\doibase
  10.1103/PhysRevLett.105.136805} {\bibfield  {journal} {\bibinfo  {journal}
  {Phys. Rev. Lett.}\ }\textbf {\bibinfo {volume} {105}},\ \bibinfo {pages}
  {136805} (\bibinfo {year} {2010})}\BibitemShut {NoStop}%
\bibitem [{\citenamefont {Xu}\ \emph {et~al.}(2014)\citenamefont {Xu},
  \citenamefont {Yao}, \citenamefont {Xiao},\ and\ \citenamefont
  {Heinz}}]{Xu_NatPhys_2014}%
  \BibitemOpen
  \bibfield  {author} {\bibinfo {author} {\bibfnamefont {X.}~\bibnamefont
  {Xu}}, \bibinfo {author} {\bibfnamefont {W.}~\bibnamefont {Yao}}, \bibinfo
  {author} {\bibfnamefont {D.}~\bibnamefont {Xiao}}, \ and\ \bibinfo {author}
  {\bibfnamefont {T.~F.}\ \bibnamefont {Heinz}},\ }\href {\doibase
  10.1038/NPHYS2942} {\bibfield  {journal} {\bibinfo  {journal} {Nat. Phys.}\
  }\textbf {\bibinfo {volume} {10}},\ \bibinfo {pages} {343} (\bibinfo {year}
  {2014})}\BibitemShut {NoStop}%
\bibitem [{\citenamefont {Manzeli}\ \emph {et~al.}(2017)\citenamefont
  {Manzeli}, \citenamefont {Ovchinnikov}, \citenamefont {Pasquier},
  \citenamefont {Yazyev},\ and\ \citenamefont
  {Kis}}]{Manzeli_NatRevMater_2017}%
  \BibitemOpen
  \bibfield  {author} {\bibinfo {author} {\bibfnamefont {S.}~\bibnamefont
  {Manzeli}}, \bibinfo {author} {\bibfnamefont {D.}~\bibnamefont
  {Ovchinnikov}}, \bibinfo {author} {\bibfnamefont {D.}~\bibnamefont
  {Pasquier}}, \bibinfo {author} {\bibfnamefont {O.~V.}\ \bibnamefont
  {Yazyev}}, \ and\ \bibinfo {author} {\bibfnamefont {A.}~\bibnamefont {Kis}},\
  }\href {\doibase 10.1038/natrevmats.2017.33} {\bibfield  {journal} {\bibinfo
  {journal} {Nat. Rev. Mater.}\ }\textbf {\bibinfo {volume} {2}},\ \bibinfo
  {pages} {17033} (\bibinfo {year} {2017})}\BibitemShut {NoStop}%
\bibitem [{\citenamefont {Mak}\ \emph {et~al.}(2018)\citenamefont {Mak},
  \citenamefont {Xiao},\ and\ \citenamefont {Shan}}]{Mak_NatPhoton_2018}%
  \BibitemOpen
  \bibfield  {author} {\bibinfo {author} {\bibfnamefont {K.~F.}\ \bibnamefont
  {Mak}}, \bibinfo {author} {\bibfnamefont {D.}~\bibnamefont {Xiao}}, \ and\
  \bibinfo {author} {\bibfnamefont {J.}~\bibnamefont {Shan}},\ }\href {\doibase
  10.1038/s41566-018-0204-6} {\bibfield  {journal} {\bibinfo  {journal} {Nat.
  Photon.}\ }\textbf {\bibinfo {volume} {12}},\ \bibinfo {pages} {451}
  (\bibinfo {year} {2018})}\BibitemShut {NoStop}%
\bibitem [{\citenamefont {Chernikov}\ \emph {et~al.}(2014)\citenamefont
  {Chernikov}, \citenamefont {Berkelbach}, \citenamefont {Hill}, \citenamefont
  {Rigosi}, \citenamefont {Li}, \citenamefont {Aslan}, \citenamefont
  {Reichman}, \citenamefont {Hybertsen},\ and\ \citenamefont
  {Heinz}}]{Chernikov_PRL_2014}%
  \BibitemOpen
  \bibfield  {author} {\bibinfo {author} {\bibfnamefont {A.}~\bibnamefont
  {Chernikov}}, \bibinfo {author} {\bibfnamefont {T.~C.}\ \bibnamefont
  {Berkelbach}}, \bibinfo {author} {\bibfnamefont {H.~M.}\ \bibnamefont
  {Hill}}, \bibinfo {author} {\bibfnamefont {A.}~\bibnamefont {Rigosi}},
  \bibinfo {author} {\bibfnamefont {Y.}~\bibnamefont {Li}}, \bibinfo {author}
  {\bibfnamefont {O.~B.}\ \bibnamefont {Aslan}}, \bibinfo {author}
  {\bibfnamefont {D.~R.}\ \bibnamefont {Reichman}}, \bibinfo {author}
  {\bibfnamefont {M.~S.}\ \bibnamefont {Hybertsen}}, \ and\ \bibinfo {author}
  {\bibfnamefont {T.~F.}\ \bibnamefont {Heinz}},\ }\href {\doibase
  10.1103/PhysRevLett.113.076802} {\bibfield  {journal} {\bibinfo  {journal}
  {Phys. Rev. Lett.}\ }\textbf {\bibinfo {volume} {113}},\ \bibinfo {pages}
  {076802} (\bibinfo {year} {2014})}\BibitemShut {NoStop}%
\bibitem [{\citenamefont {Robert}\ \emph {et~al.}(2018)\citenamefont {Robert},
  \citenamefont {Semina}, \citenamefont {Cadiz}, \citenamefont {Manca},
  \citenamefont {Courtade}, \citenamefont {Taniguchi}, \citenamefont
  {Watanabe}, \citenamefont {Cai}, \citenamefont {Tongay}, \citenamefont
  {Lassagne}, \citenamefont {Renucci}, \citenamefont {Amand}, \citenamefont
  {Marie}, \citenamefont {Glazov},\ and\ \citenamefont
  {Urbaszek}}]{Robert_PRM_2018}%
  \BibitemOpen
  \bibfield  {author} {\bibinfo {author} {\bibfnamefont {C.}~\bibnamefont
  {Robert}}, \bibinfo {author} {\bibfnamefont {M.~A.}\ \bibnamefont {Semina}},
  \bibinfo {author} {\bibfnamefont {F.}~\bibnamefont {Cadiz}}, \bibinfo
  {author} {\bibfnamefont {M.}~\bibnamefont {Manca}}, \bibinfo {author}
  {\bibfnamefont {E.}~\bibnamefont {Courtade}}, \bibinfo {author}
  {\bibfnamefont {T.}~\bibnamefont {Taniguchi}}, \bibinfo {author}
  {\bibfnamefont {K.}~\bibnamefont {Watanabe}}, \bibinfo {author}
  {\bibfnamefont {H.}~\bibnamefont {Cai}}, \bibinfo {author} {\bibfnamefont
  {S.}~\bibnamefont {Tongay}}, \bibinfo {author} {\bibfnamefont
  {B.}~\bibnamefont {Lassagne}}, \bibinfo {author} {\bibfnamefont
  {P.}~\bibnamefont {Renucci}}, \bibinfo {author} {\bibfnamefont
  {T.}~\bibnamefont {Amand}}, \bibinfo {author} {\bibfnamefont
  {X.}~\bibnamefont {Marie}}, \bibinfo {author} {\bibfnamefont {M.~M.}\
  \bibnamefont {Glazov}}, \ and\ \bibinfo {author} {\bibfnamefont
  {B.}~\bibnamefont {Urbaszek}},\ }\href {\doibase
  10.1103/PhysRevMaterials.2.011001} {\bibfield  {journal} {\bibinfo  {journal}
  {Phys. Rev. Materials}\ }\textbf {\bibinfo {volume} {2}},\ \bibinfo {pages}
  {011001} (\bibinfo {year} {2018})}\BibitemShut {NoStop}%
\bibitem [{\citenamefont {Stier}\ \emph {et~al.}(2018)\citenamefont {Stier},
  \citenamefont {Wilson}, \citenamefont {Velizhanin}, \citenamefont {Kono},
  \citenamefont {Xu},\ and\ \citenamefont {Crooker}}]{Stier_PRL_2018}%
  \BibitemOpen
  \bibfield  {author} {\bibinfo {author} {\bibfnamefont {A.~V.}\ \bibnamefont
  {Stier}}, \bibinfo {author} {\bibfnamefont {N.~P.}\ \bibnamefont {Wilson}},
  \bibinfo {author} {\bibfnamefont {K.~A.}\ \bibnamefont {Velizhanin}},
  \bibinfo {author} {\bibfnamefont {J.}~\bibnamefont {Kono}}, \bibinfo {author}
  {\bibfnamefont {X.}~\bibnamefont {Xu}}, \ and\ \bibinfo {author}
  {\bibfnamefont {S.~A.}\ \bibnamefont {Crooker}},\ }\href {\doibase
  10.1103/PhysRevLett.120.057405} {\bibfield  {journal} {\bibinfo  {journal}
  {Phys. Rev. Lett.}\ }\textbf {\bibinfo {volume} {120}},\ \bibinfo {pages}
  {057405} (\bibinfo {year} {2018})}\BibitemShut {NoStop}%
\bibitem [{\citenamefont {Xiao}\ \emph {et~al.}(2012)\citenamefont {Xiao},
  \citenamefont {Liu}, \citenamefont {Feng}, \citenamefont {Xu},\ and\
  \citenamefont {Yao}}]{Xiao_PRL_2012}%
  \BibitemOpen
  \bibfield  {author} {\bibinfo {author} {\bibfnamefont {D.}~\bibnamefont
  {Xiao}}, \bibinfo {author} {\bibfnamefont {G.-B.}\ \bibnamefont {Liu}},
  \bibinfo {author} {\bibfnamefont {W.}~\bibnamefont {Feng}}, \bibinfo {author}
  {\bibfnamefont {X.}~\bibnamefont {Xu}}, \ and\ \bibinfo {author}
  {\bibfnamefont {W.}~\bibnamefont {Yao}},\ }\href {\doibase
  10.1103/PhysRevLett.108.196802} {\bibfield  {journal} {\bibinfo  {journal}
  {Phys. Rev. Lett.}\ }\textbf {\bibinfo {volume} {108}},\ \bibinfo {pages}
  {196802} (\bibinfo {year} {2012})}\BibitemShut {NoStop}%
\bibitem [{\citenamefont {Cai}\ \emph {et~al.}(2013)\citenamefont {Cai},
  \citenamefont {Yang}, \citenamefont {Li}, \citenamefont {Zhang},
  \citenamefont {Shi}, \citenamefont {Yao},\ and\ \citenamefont
  {Niu}}]{Cai_PRB_2013}%
  \BibitemOpen
  \bibfield  {author} {\bibinfo {author} {\bibfnamefont {T.}~\bibnamefont
  {Cai}}, \bibinfo {author} {\bibfnamefont {S.~A.}\ \bibnamefont {Yang}},
  \bibinfo {author} {\bibfnamefont {X.}~\bibnamefont {Li}}, \bibinfo {author}
  {\bibfnamefont {F.}~\bibnamefont {Zhang}}, \bibinfo {author} {\bibfnamefont
  {J.}~\bibnamefont {Shi}}, \bibinfo {author} {\bibfnamefont {W.}~\bibnamefont
  {Yao}}, \ and\ \bibinfo {author} {\bibfnamefont {Q.}~\bibnamefont {Niu}},\
  }\href {\doibase 10.1103/PhysRevB.88.115140} {\bibfield  {journal} {\bibinfo
  {journal} {Phys. Rev. B}\ }\textbf {\bibinfo {volume} {88}},\ \bibinfo
  {pages} {115140} (\bibinfo {year} {2013})}\BibitemShut {NoStop}%
\bibitem [{\citenamefont {Rose}\ \emph {et~al.}(2013)\citenamefont {Rose},
  \citenamefont {Goerbig},\ and\ \citenamefont {Pi\'echon}}]{Rose_PRB_2013}%
  \BibitemOpen
  \bibfield  {author} {\bibinfo {author} {\bibfnamefont {F.}~\bibnamefont
  {Rose}}, \bibinfo {author} {\bibfnamefont {M.~O.}\ \bibnamefont {Goerbig}}, \
  and\ \bibinfo {author} {\bibfnamefont {F.}~\bibnamefont {Pi\'echon}},\ }\href
  {\doibase 10.1103/PhysRevB.88.125438} {\bibfield  {journal} {\bibinfo
  {journal} {Phys. Rev. B}\ }\textbf {\bibinfo {volume} {88}},\ \bibinfo
  {pages} {125438} (\bibinfo {year} {2013})}\BibitemShut {NoStop}%
\bibitem [{\citenamefont {Srivastava}\ \emph {et~al.}(2015)\citenamefont
  {Srivastava}, \citenamefont {Sidler}, \citenamefont {Allain}, \citenamefont
  {Lembke}, \citenamefont {Kis},\ and\ \citenamefont
  {Imamoglu}}]{Srivastava_NP_2015}%
  \BibitemOpen
  \bibfield  {author} {\bibinfo {author} {\bibfnamefont {A.}~\bibnamefont
  {Srivastava}}, \bibinfo {author} {\bibfnamefont {M.}~\bibnamefont {Sidler}},
  \bibinfo {author} {\bibfnamefont {A.~V.}\ \bibnamefont {Allain}}, \bibinfo
  {author} {\bibfnamefont {D.~S.}\ \bibnamefont {Lembke}}, \bibinfo {author}
  {\bibfnamefont {A.}~\bibnamefont {Kis}}, \ and\ \bibinfo {author}
  {\bibfnamefont {A.}~\bibnamefont {Imamoglu}},\ }\href {\doibase
  10.1038/NPHYS3203} {\bibfield  {journal} {\bibinfo  {journal} {Nat. Phys.}\
  }\textbf {\bibinfo {volume} {11}},\ \bibinfo {pages} {141} (\bibinfo {year}
  {2015})}\BibitemShut {NoStop}%
\bibitem [{\citenamefont {Aivazian}\ \emph {et~al.}(2015)\citenamefont
  {Aivazian}, \citenamefont {Gong}, \citenamefont {Jones}, \citenamefont {Chu},
  \citenamefont {Yan}, \citenamefont {Mandrus}, \citenamefont {Zhang},
  \citenamefont {Cobden}, \citenamefont {Yao},\ and\ \citenamefont
  {Xu}}]{Aivazian_NP_2015}%
  \BibitemOpen
  \bibfield  {author} {\bibinfo {author} {\bibfnamefont {G.}~\bibnamefont
  {Aivazian}}, \bibinfo {author} {\bibfnamefont {Z.}~\bibnamefont {Gong}},
  \bibinfo {author} {\bibfnamefont {A.~M.}\ \bibnamefont {Jones}}, \bibinfo
  {author} {\bibfnamefont {R.-L.}\ \bibnamefont {Chu}}, \bibinfo {author}
  {\bibfnamefont {J.}~\bibnamefont {Yan}}, \bibinfo {author} {\bibfnamefont
  {D.~G.}\ \bibnamefont {Mandrus}}, \bibinfo {author} {\bibfnamefont
  {C.}~\bibnamefont {Zhang}}, \bibinfo {author} {\bibfnamefont
  {D.}~\bibnamefont {Cobden}}, \bibinfo {author} {\bibfnamefont
  {W.}~\bibnamefont {Yao}}, \ and\ \bibinfo {author} {\bibfnamefont
  {X.}~\bibnamefont {Xu}},\ }\href {\doibase 10.1038/NPHYS3201} {\bibfield
  {journal} {\bibinfo  {journal} {Nat. Phys.}\ }\textbf {\bibinfo {volume}
  {11}},\ \bibinfo {pages} {148} (\bibinfo {year} {2015})}\BibitemShut
  {NoStop}%
\bibitem [{\citenamefont {Fallahazad}\ \emph {et~al.}(2016)\citenamefont
  {Fallahazad}, \citenamefont {Movva}, \citenamefont {Kim}, \citenamefont
  {Larentis}, \citenamefont {Taniguchi}, \citenamefont {Watanabe},
  \citenamefont {Banerjee},\ and\ \citenamefont {Tutuc}}]{Fallahazad_PRL_2016}%
  \BibitemOpen
  \bibfield  {author} {\bibinfo {author} {\bibfnamefont {B.}~\bibnamefont
  {Fallahazad}}, \bibinfo {author} {\bibfnamefont {H.~C.~P.}\ \bibnamefont
  {Movva}}, \bibinfo {author} {\bibfnamefont {K.}~\bibnamefont {Kim}}, \bibinfo
  {author} {\bibfnamefont {S.}~\bibnamefont {Larentis}}, \bibinfo {author}
  {\bibfnamefont {T.}~\bibnamefont {Taniguchi}}, \bibinfo {author}
  {\bibfnamefont {K.}~\bibnamefont {Watanabe}}, \bibinfo {author}
  {\bibfnamefont {S.~K.}\ \bibnamefont {Banerjee}}, \ and\ \bibinfo {author}
  {\bibfnamefont {E.}~\bibnamefont {Tutuc}},\ }\href {\doibase
  10.1103/PhysRevLett.116.086601} {\bibfield  {journal} {\bibinfo  {journal}
  {Phys. Rev. Lett.}\ }\textbf {\bibinfo {volume} {116}},\ \bibinfo {pages}
  {086601} (\bibinfo {year} {2016})}\BibitemShut {NoStop}%
\bibitem [{\citenamefont {Movva}\ \emph {et~al.}(2017)\citenamefont {Movva},
  \citenamefont {Fallahazad}, \citenamefont {Kim}, \citenamefont {Larentis},
  \citenamefont {Taniguchi}, \citenamefont {Watanabe}, \citenamefont
  {Banerjee},\ and\ \citenamefont {Tutuc}}]{Movva_PRL_2017}%
  \BibitemOpen
  \bibfield  {author} {\bibinfo {author} {\bibfnamefont {H.~C.~P.}\
  \bibnamefont {Movva}}, \bibinfo {author} {\bibfnamefont {B.}~\bibnamefont
  {Fallahazad}}, \bibinfo {author} {\bibfnamefont {K.}~\bibnamefont {Kim}},
  \bibinfo {author} {\bibfnamefont {S.}~\bibnamefont {Larentis}}, \bibinfo
  {author} {\bibfnamefont {T.}~\bibnamefont {Taniguchi}}, \bibinfo {author}
  {\bibfnamefont {K.}~\bibnamefont {Watanabe}}, \bibinfo {author}
  {\bibfnamefont {S.~K.}\ \bibnamefont {Banerjee}}, \ and\ \bibinfo {author}
  {\bibfnamefont {E.}~\bibnamefont {Tutuc}},\ }\href {\doibase
  10.1103/PhysRevLett.118.247701} {\bibfield  {journal} {\bibinfo  {journal}
  {Phys. Rev. Lett.}\ }\textbf {\bibinfo {volume} {118}},\ \bibinfo {pages}
  {247701} (\bibinfo {year} {2017})}\BibitemShut {NoStop}%
\bibitem [{\citenamefont {Gustafsson}\ \emph {et~al.}(2018)\citenamefont
  {Gustafsson}, \citenamefont {Yankowitz}, \citenamefont {Forsythe},
  \citenamefont {Rhodes}, \citenamefont {Watanabe}, \citenamefont {Taniguchi},
  \citenamefont {Hone}, \citenamefont {Zhu},\ and\ \citenamefont
  {Dean}}]{Gustafsson_NatMater_2018}%
  \BibitemOpen
  \bibfield  {author} {\bibinfo {author} {\bibfnamefont {M.~V.}\ \bibnamefont
  {Gustafsson}}, \bibinfo {author} {\bibfnamefont {M.}~\bibnamefont
  {Yankowitz}}, \bibinfo {author} {\bibfnamefont {C.}~\bibnamefont {Forsythe}},
  \bibinfo {author} {\bibfnamefont {D.}~\bibnamefont {Rhodes}}, \bibinfo
  {author} {\bibfnamefont {K.}~\bibnamefont {Watanabe}}, \bibinfo {author}
  {\bibfnamefont {T.}~\bibnamefont {Taniguchi}}, \bibinfo {author}
  {\bibfnamefont {J.}~\bibnamefont {Hone}}, \bibinfo {author} {\bibfnamefont
  {X.}~\bibnamefont {Zhu}}, \ and\ \bibinfo {author} {\bibfnamefont {C.~R.}\
  \bibnamefont {Dean}},\ }\href {\doibase 10.1038/s41563-018-0036-2} {\bibfield
   {journal} {\bibinfo  {journal} {Nat. Mater.}\ }\textbf {\bibinfo {volume}
  {17}},\ \bibinfo {pages} {411} (\bibinfo {year} {2018})}\BibitemShut
  {NoStop}%
\bibitem [{\citenamefont {Larentis}\ \emph {et~al.}(2018)\citenamefont
  {Larentis}, \citenamefont {Movva}, \citenamefont {Fallahazad}, \citenamefont
  {Kim}, \citenamefont {Behroozi}, \citenamefont {Taniguchi}, \citenamefont
  {Watanabe}, \citenamefont {Banerjee},\ and\ \citenamefont
  {Tutuc}}]{Larentis_PRB_2018}%
  \BibitemOpen
  \bibfield  {author} {\bibinfo {author} {\bibfnamefont {S.}~\bibnamefont
  {Larentis}}, \bibinfo {author} {\bibfnamefont {H.~C.~P.}\ \bibnamefont
  {Movva}}, \bibinfo {author} {\bibfnamefont {B.}~\bibnamefont {Fallahazad}},
  \bibinfo {author} {\bibfnamefont {K.}~\bibnamefont {Kim}}, \bibinfo {author}
  {\bibfnamefont {A.}~\bibnamefont {Behroozi}}, \bibinfo {author}
  {\bibfnamefont {T.}~\bibnamefont {Taniguchi}}, \bibinfo {author}
  {\bibfnamefont {K.}~\bibnamefont {Watanabe}}, \bibinfo {author}
  {\bibfnamefont {S.~K.}\ \bibnamefont {Banerjee}}, \ and\ \bibinfo {author}
  {\bibfnamefont {E.}~\bibnamefont {Tutuc}},\ }\href {\doibase
  10.1103/PhysRevB.97.201407} {\bibfield  {journal} {\bibinfo  {journal} {Phys.
  Rev. B}\ }\textbf {\bibinfo {volume} {97}},\ \bibinfo {pages} {201407}
  (\bibinfo {year} {2018})}\BibitemShut {NoStop}%
\bibitem [{\citenamefont {Lin}\ \emph {et~al.}(2018)\citenamefont {Lin},
  \citenamefont {Han}, \citenamefont {Piot}, \citenamefont {Wu}, \citenamefont
  {Xu}, \citenamefont {Long}, \citenamefont {An}, \citenamefont
  {Ka~Man~Cheung}, \citenamefont {Zheng}, \citenamefont {Plochocka},
  \citenamefont {Maude}, \citenamefont {Zhang},\ and\ \citenamefont
  {Wang}}]{Lin_arXiv_2018}%
  \BibitemOpen
  \bibfield  {author} {\bibinfo {author} {\bibfnamefont {J.}~\bibnamefont
  {Lin}}, \bibinfo {author} {\bibfnamefont {T.}~\bibnamefont {Han}}, \bibinfo
  {author} {\bibfnamefont {B.~A.}\ \bibnamefont {Piot}}, \bibinfo {author}
  {\bibfnamefont {Z.}~\bibnamefont {Wu}}, \bibinfo {author} {\bibfnamefont
  {S.~X.}\ \bibnamefont {Xu}}, \bibinfo {author} {\bibfnamefont
  {G.}~\bibnamefont {Long}}, \bibinfo {author} {\bibfnamefont {L.}~\bibnamefont
  {An}}, \bibinfo {author} {\bibfnamefont {P.}~\bibnamefont {Ka~Man~Cheung}},
  \bibinfo {author} {\bibfnamefont {P.-P.}\ \bibnamefont {Zheng}}, \bibinfo
  {author} {\bibfnamefont {P.}~\bibnamefont {Plochocka}}, \bibinfo {author}
  {\bibfnamefont {D.~K.}\ \bibnamefont {Maude}}, \bibinfo {author}
  {\bibfnamefont {F.}~\bibnamefont {Zhang}}, \ and\ \bibinfo {author}
  {\bibfnamefont {N.}~\bibnamefont {Wang}},\ }\href
  {https://arxiv.org/abs/1803.08007} {\bibfield  {journal} {\bibinfo  {journal}
  {arXiv:1803.08007}\ } (\bibinfo {year} {2018})}\BibitemShut {NoStop}%
\bibitem [{\citenamefont {Pisoni}\ \emph {et~al.}(2018)\citenamefont {Pisoni},
  \citenamefont {Kormányos}, \citenamefont {Brooks}, \citenamefont {Lei},
  \citenamefont {Back}, \citenamefont {Eich}, \citenamefont {Overweg},
  \citenamefont {Lee}, \citenamefont {Rickhaus}, \citenamefont {Watanabe},
  \citenamefont {Taniguchi}, \citenamefont {Imamoglu}, \citenamefont {Burkard},
  \citenamefont {Ihn},\ and\ \citenamefont {Ensslin}}]{Pisoni_arXiv_2018}%
  \BibitemOpen
  \bibfield  {author} {\bibinfo {author} {\bibfnamefont {R.}~\bibnamefont
  {Pisoni}}, \bibinfo {author} {\bibfnamefont {A.}~\bibnamefont {Kormányos}},
  \bibinfo {author} {\bibfnamefont {M.}~\bibnamefont {Brooks}}, \bibinfo
  {author} {\bibfnamefont {Z.}~\bibnamefont {Lei}}, \bibinfo {author}
  {\bibfnamefont {P.}~\bibnamefont {Back}}, \bibinfo {author} {\bibfnamefont
  {M.}~\bibnamefont {Eich}}, \bibinfo {author} {\bibfnamefont {H.}~\bibnamefont
  {Overweg}}, \bibinfo {author} {\bibfnamefont {Y.}~\bibnamefont {Lee}},
  \bibinfo {author} {\bibfnamefont {P.}~\bibnamefont {Rickhaus}}, \bibinfo
  {author} {\bibfnamefont {K.}~\bibnamefont {Watanabe}}, \bibinfo {author}
  {\bibfnamefont {T.}~\bibnamefont {Taniguchi}}, \bibinfo {author}
  {\bibfnamefont {A.}~\bibnamefont {Imamoglu}}, \bibinfo {author}
  {\bibfnamefont {G.}~\bibnamefont {Burkard}}, \bibinfo {author} {\bibfnamefont
  {T.}~\bibnamefont {Ihn}}, \ and\ \bibinfo {author} {\bibfnamefont
  {K.}~\bibnamefont {Ensslin}},\ }\href {https://arxiv.org/abs/1806.06402}
  {\bibfield  {journal} {\bibinfo  {journal} {arXiv:1806.06402}\ } (\bibinfo
  {year} {2018})}\BibitemShut {NoStop}%
\bibitem [{\citenamefont {Wang}\ \emph {et~al.}(2017)\citenamefont {Wang},
  \citenamefont {Shan},\ and\ \citenamefont {Mak}}]{Wang_NatNano_2017}%
  \BibitemOpen
  \bibfield  {author} {\bibinfo {author} {\bibfnamefont {Z.}~\bibnamefont
  {Wang}}, \bibinfo {author} {\bibfnamefont {J.}~\bibnamefont {Shan}}, \ and\
  \bibinfo {author} {\bibfnamefont {K.~F.}\ \bibnamefont {Mak}},\ }\href
  {\doibase 10.1038/NNANO.2016.213} {\bibfield  {journal} {\bibinfo  {journal}
  {Nat. Nanotechnol.}\ }\textbf {\bibinfo {volume} {12}},\ \bibinfo {pages}
  {144} (\bibinfo {year} {2017})}\BibitemShut {NoStop}%
\bibitem [{\citenamefont {Sidler}\ \emph {et~al.}(2017)\citenamefont {Sidler},
  \citenamefont {Back}, \citenamefont {Cotlet}, \citenamefont {Srivastava},
  \citenamefont {Fink}, \citenamefont {Kroner}, \citenamefont {Demler},\ and\
  \citenamefont {Imamoglu}}]{Sidler_NatPhys_2017}%
  \BibitemOpen
  \bibfield  {author} {\bibinfo {author} {\bibfnamefont {M.}~\bibnamefont
  {Sidler}}, \bibinfo {author} {\bibfnamefont {P.}~\bibnamefont {Back}},
  \bibinfo {author} {\bibfnamefont {O.}~\bibnamefont {Cotlet}}, \bibinfo
  {author} {\bibfnamefont {A.}~\bibnamefont {Srivastava}}, \bibinfo {author}
  {\bibfnamefont {T.}~\bibnamefont {Fink}}, \bibinfo {author} {\bibfnamefont
  {M.}~\bibnamefont {Kroner}}, \bibinfo {author} {\bibfnamefont
  {E.}~\bibnamefont {Demler}}, \ and\ \bibinfo {author} {\bibfnamefont
  {A.}~\bibnamefont {Imamoglu}},\ }\href {\doibase 10.1038/nphys3949}
  {\bibfield  {journal} {\bibinfo  {journal} {Nat. Phys.}\ }\textbf {\bibinfo
  {volume} {13}},\ \bibinfo {pages} {255} (\bibinfo {year} {2017})}\BibitemShut
  {NoStop}%
\bibitem [{\citenamefont {Efimkin}\ and\ \citenamefont
  {MacDonald}(2017)}]{Efimkin_PRB_2017}%
  \BibitemOpen
  \bibfield  {author} {\bibinfo {author} {\bibfnamefont {D.~K.}\ \bibnamefont
  {Efimkin}}\ and\ \bibinfo {author} {\bibfnamefont {A.~H.}\ \bibnamefont
  {MacDonald}},\ }\href {\doibase 10.1103/PhysRevB.95.035417} {\bibfield
  {journal} {\bibinfo  {journal} {Phys. Rev. B}\ }\textbf {\bibinfo {volume}
  {95}},\ \bibinfo {pages} {035417} (\bibinfo {year} {2017})}\BibitemShut
  {NoStop}%
\bibitem [{\citenamefont {Back}\ \emph {et~al.}(2018)\citenamefont {Back},
  \citenamefont {Zeytinoglu}, \citenamefont {Ijaz}, \citenamefont {Kroner},\
  and\ \citenamefont {Imamo\ifmmode~\breve{g}\else
  \u{g}\fi{}lu}}]{Back_PRL_2018}%
  \BibitemOpen
  \bibfield  {author} {\bibinfo {author} {\bibfnamefont {P.}~\bibnamefont
  {Back}}, \bibinfo {author} {\bibfnamefont {S.}~\bibnamefont {Zeytinoglu}},
  \bibinfo {author} {\bibfnamefont {A.}~\bibnamefont {Ijaz}}, \bibinfo {author}
  {\bibfnamefont {M.}~\bibnamefont {Kroner}}, \ and\ \bibinfo {author}
  {\bibfnamefont {A.}~\bibnamefont {Imamo\ifmmode~\breve{g}\else
  \u{g}\fi{}lu}},\ }\href {\doibase 10.1103/PhysRevLett.120.037401} {\bibfield
  {journal} {\bibinfo  {journal} {Phys. Rev. Lett.}\ }\textbf {\bibinfo
  {volume} {120}},\ \bibinfo {pages} {037401} (\bibinfo {year}
  {2018})}\BibitemShut {NoStop}%
\bibitem [{\citenamefont {Efimkin}\ and\ \citenamefont
  {MacDonald}(2018)}]{Efimkin_PRB_2018}%
  \BibitemOpen
  \bibfield  {author} {\bibinfo {author} {\bibfnamefont {D.~K.}\ \bibnamefont
  {Efimkin}}\ and\ \bibinfo {author} {\bibfnamefont {A.~H.}\ \bibnamefont
  {MacDonald}},\ }\href {\doibase 10.1103/PhysRevB.97.235432} {\bibfield
  {journal} {\bibinfo  {journal} {Phys. Rev. B}\ }\textbf {\bibinfo {volume}
  {97}},\ \bibinfo {pages} {235432} (\bibinfo {year} {2018})}\BibitemShut
  {NoStop}%
\bibitem [{SM()}]{SM}%
  \BibitemOpen
  \href@noop {} {}\bibinfo {note} {See Supplemental Material at [URL will be
  inserted by publisher] for further information concerning data analysis,
  sample fabrication, experimental setup, calibration of the hole doping
  density, reproducibility of the results on other devices, and description of
  the theoretical model.}\BibitemShut {Stop}%
\bibitem [{\citenamefont {Back}\ \emph {et~al.}(2017)\citenamefont {Back},
  \citenamefont {Sidler}, \citenamefont {Cotlet}, \citenamefont {Srivastava},
  \citenamefont {Takemura}, \citenamefont {Kroner},\ and\ \citenamefont
  {Imamo\ifmmode~\breve{g}\else \u{g}\fi{}lu}}]{Back_PRL_2017}%
  \BibitemOpen
  \bibfield  {author} {\bibinfo {author} {\bibfnamefont {P.}~\bibnamefont
  {Back}}, \bibinfo {author} {\bibfnamefont {M.}~\bibnamefont {Sidler}},
  \bibinfo {author} {\bibfnamefont {O.}~\bibnamefont {Cotlet}}, \bibinfo
  {author} {\bibfnamefont {A.}~\bibnamefont {Srivastava}}, \bibinfo {author}
  {\bibfnamefont {N.}~\bibnamefont {Takemura}}, \bibinfo {author}
  {\bibfnamefont {M.}~\bibnamefont {Kroner}}, \ and\ \bibinfo {author}
  {\bibfnamefont {A.}~\bibnamefont {Imamo\ifmmode~\breve{g}\else
  \u{g}\fi{}lu}},\ }\href {\doibase 10.1103/PhysRevLett.118.237404} {\bibfield
  {journal} {\bibinfo  {journal} {Phys. Rev. Lett.}\ }\textbf {\bibinfo
  {volume} {118}},\ \bibinfo {pages} {237404} (\bibinfo {year}
  {2017})}\BibitemShut {NoStop}%
\bibitem [{\citenamefont {Cheiwchanchamnangij}\ and\ \citenamefont
  {Lambrecht}(2012)}]{Cheiwchanchamnangij_PRB_2012}%
  \BibitemOpen
  \bibfield  {author} {\bibinfo {author} {\bibfnamefont {T.}~\bibnamefont
  {Cheiwchanchamnangij}}\ and\ \bibinfo {author} {\bibfnamefont {W.~R.~L.}\
  \bibnamefont {Lambrecht}},\ }\href {\doibase 10.1103/PhysRevB.85.205302}
  {\bibfield  {journal} {\bibinfo  {journal} {Phys. Rev. B}\ }\textbf {\bibinfo
  {volume} {85}},\ \bibinfo {pages} {205302} (\bibinfo {year}
  {2012})}\BibitemShut {NoStop}%
\bibitem [{\citenamefont {Yun}\ \emph {et~al.}(2012)\citenamefont {Yun},
  \citenamefont {Han}, \citenamefont {Hong}, \citenamefont {Kim},\ and\
  \citenamefont {Lee}}]{Yun_PRB_2012}%
  \BibitemOpen
  \bibfield  {author} {\bibinfo {author} {\bibfnamefont {W.~S.}\ \bibnamefont
  {Yun}}, \bibinfo {author} {\bibfnamefont {S.~W.}\ \bibnamefont {Han}},
  \bibinfo {author} {\bibfnamefont {S.~C.}\ \bibnamefont {Hong}}, \bibinfo
  {author} {\bibfnamefont {I.~G.}\ \bibnamefont {Kim}}, \ and\ \bibinfo
  {author} {\bibfnamefont {J.~D.}\ \bibnamefont {Lee}},\ }\href {\doibase
  10.1103/PhysRevB.85.033305} {\bibfield  {journal} {\bibinfo  {journal} {Phys.
  Rev. B}\ }\textbf {\bibinfo {volume} {85}},\ \bibinfo {pages} {033305}
  (\bibinfo {year} {2012})}\BibitemShut {NoStop}%
\bibitem [{\citenamefont {Kormanyos}\ \emph {et~al.}(2015)\citenamefont
  {Kormanyos}, \citenamefont {Burkard}, \citenamefont {Gmitra}, \citenamefont
  {Fabian}, \citenamefont {Zolyomi}, \citenamefont {Drummond},\ and\
  \citenamefont {Falko}}]{Kormanyos_2DMat_2015}%
  \BibitemOpen
  \bibfield  {author} {\bibinfo {author} {\bibfnamefont {A.}~\bibnamefont
  {Kormanyos}}, \bibinfo {author} {\bibfnamefont {G.}~\bibnamefont {Burkard}},
  \bibinfo {author} {\bibfnamefont {M.}~\bibnamefont {Gmitra}}, \bibinfo
  {author} {\bibfnamefont {J.}~\bibnamefont {Fabian}}, \bibinfo {author}
  {\bibfnamefont {V.}~\bibnamefont {Zolyomi}}, \bibinfo {author} {\bibfnamefont
  {N.~D.}\ \bibnamefont {Drummond}}, \ and\ \bibinfo {author} {\bibfnamefont
  {V.}~\bibnamefont {Falko}},\ }\href {\doibase 10.1088/2053-1583/2/2/022001}
  {\bibfield  {journal} {\bibinfo  {journal} {2D Mater.}\ }\textbf {\bibinfo
  {volume} {022001}},\ \bibinfo {pages} {6} (\bibinfo {year}
  {2015})}\BibitemShut {NoStop}%
\bibitem [{\citenamefont {Jin}\ \emph {et~al.}(2015)\citenamefont {Jin},
  \citenamefont {Yeh}, \citenamefont {Zaki}, \citenamefont {Zhang},
  \citenamefont {Liou}, \citenamefont {Sadowski}, \citenamefont {Barinov},
  \citenamefont {Yablonskikh}, \citenamefont {Dadap}, \citenamefont {Sutter},
  \citenamefont {Herman},\ and\ \citenamefont {Osgood}}]{Jin_PRB_2015}%
  \BibitemOpen
  \bibfield  {author} {\bibinfo {author} {\bibfnamefont {W.}~\bibnamefont
  {Jin}}, \bibinfo {author} {\bibfnamefont {P.-C.}\ \bibnamefont {Yeh}},
  \bibinfo {author} {\bibfnamefont {N.}~\bibnamefont {Zaki}}, \bibinfo {author}
  {\bibfnamefont {D.}~\bibnamefont {Zhang}}, \bibinfo {author} {\bibfnamefont
  {J.~T.}\ \bibnamefont {Liou}}, \bibinfo {author} {\bibfnamefont {J.~T.}\
  \bibnamefont {Sadowski}}, \bibinfo {author} {\bibfnamefont {A.}~\bibnamefont
  {Barinov}}, \bibinfo {author} {\bibfnamefont {M.}~\bibnamefont
  {Yablonskikh}}, \bibinfo {author} {\bibfnamefont {J.~I.}\ \bibnamefont
  {Dadap}}, \bibinfo {author} {\bibfnamefont {P.}~\bibnamefont {Sutter}},
  \bibinfo {author} {\bibfnamefont {I.~P.}\ \bibnamefont {Herman}}, \ and\
  \bibinfo {author} {\bibfnamefont {R.~M.}\ \bibnamefont {Osgood}},\ }\href
  {\doibase 10.1103/PhysRevB.91.121409} {\bibfield  {journal} {\bibinfo
  {journal} {Phys. Rev. B}\ }\textbf {\bibinfo {volume} {91}},\ \bibinfo
  {pages} {121409} (\bibinfo {year} {2015})}\BibitemShut {NoStop}%
\bibitem [{\citenamefont {Tanabe}\ \emph {et~al.}(2016)\citenamefont {Tanabe},
  \citenamefont {Gomez}, \citenamefont {Coley}, \citenamefont {Le},
  \citenamefont {Echeverria}, \citenamefont {Stecklein}, \citenamefont
  {Kandyba}, \citenamefont {Balijepalli}, \citenamefont {Klee}, \citenamefont
  {Nguyen}, \citenamefont {Preciado}, \citenamefont {Lu}, \citenamefont
  {Bobek}, \citenamefont {Barroso}, \citenamefont {{Martinez-Ta}},
  \citenamefont {Barinov}, \citenamefont {Rahman}, \citenamefont {Dowben},
  \citenamefont {Crowell},\ and\ \citenamefont {Bartels}}]{Tanabe_APL_2016}%
  \BibitemOpen
  \bibfield  {author} {\bibinfo {author} {\bibfnamefont {I.}~\bibnamefont
  {Tanabe}}, \bibinfo {author} {\bibfnamefont {M.}~\bibnamefont {Gomez}},
  \bibinfo {author} {\bibfnamefont {W.~C.}\ \bibnamefont {Coley}}, \bibinfo
  {author} {\bibfnamefont {D.}~\bibnamefont {Le}}, \bibinfo {author}
  {\bibfnamefont {E.~M.}\ \bibnamefont {Echeverria}}, \bibinfo {author}
  {\bibfnamefont {G.}~\bibnamefont {Stecklein}}, \bibinfo {author}
  {\bibfnamefont {V.}~\bibnamefont {Kandyba}}, \bibinfo {author} {\bibfnamefont
  {S.~K.}\ \bibnamefont {Balijepalli}}, \bibinfo {author} {\bibfnamefont
  {V.}~\bibnamefont {Klee}}, \bibinfo {author} {\bibfnamefont {A.~E.}\
  \bibnamefont {Nguyen}}, \bibinfo {author} {\bibfnamefont {E.}~\bibnamefont
  {Preciado}}, \bibinfo {author} {\bibfnamefont {I.}~\bibnamefont {Lu}},
  \bibinfo {author} {\bibfnamefont {S.}~\bibnamefont {Bobek}}, \bibinfo
  {author} {\bibfnamefont {D.}~\bibnamefont {Barroso}}, \bibinfo {author}
  {\bibfnamefont {D.}~\bibnamefont {{Martinez-Ta}}}, \bibinfo {author}
  {\bibfnamefont {A.}~\bibnamefont {Barinov}}, \bibinfo {author} {\bibfnamefont
  {T.~S.}\ \bibnamefont {Rahman}}, \bibinfo {author} {\bibfnamefont {P.~A.}\
  \bibnamefont {Dowben}}, \bibinfo {author} {\bibfnamefont {P.~A.}\
  \bibnamefont {Crowell}}, \ and\ \bibinfo {author} {\bibfnamefont
  {L.}~\bibnamefont {Bartels}},\ }\href {\doibase 10.1063/1.4954278} {\bibfield
   {journal} {\bibinfo  {journal} {Appl. Phys. Lett.}\ }\textbf {\bibinfo
  {volume} {108}},\ \bibinfo {pages} {252103} (\bibinfo {year}
  {2016})}\BibitemShut {NoStop}%
\bibitem [{\citenamefont {Ulstrup}\ \emph {et~al.}(2016)\citenamefont
  {Ulstrup}, \citenamefont {Katoch}, \citenamefont {Koch}, \citenamefont
  {Schwarz}, \citenamefont {Singh}, \citenamefont {{McCreary}}, \citenamefont
  {Yoo}, \citenamefont {Xu}, \citenamefont {Jonker}, \citenamefont {Kawakami},
  \citenamefont {Bostwick}, \citenamefont {Rotenberg},\ and\ \citenamefont
  {Jozwiak}}]{Ulstrup_ACSNano_2016}%
  \BibitemOpen
  \bibfield  {author} {\bibinfo {author} {\bibfnamefont {S.}~\bibnamefont
  {Ulstrup}}, \bibinfo {author} {\bibfnamefont {J.}~\bibnamefont {Katoch}},
  \bibinfo {author} {\bibfnamefont {R.~J.}\ \bibnamefont {Koch}}, \bibinfo
  {author} {\bibfnamefont {D.}~\bibnamefont {Schwarz}}, \bibinfo {author}
  {\bibfnamefont {S.}~\bibnamefont {Singh}}, \bibinfo {author} {\bibfnamefont
  {K.~M.}\ \bibnamefont {{McCreary}}}, \bibinfo {author} {\bibfnamefont
  {H.~K.}\ \bibnamefont {Yoo}}, \bibinfo {author} {\bibfnamefont
  {J.}~\bibnamefont {Xu}}, \bibinfo {author} {\bibfnamefont {B.~T.}\
  \bibnamefont {Jonker}}, \bibinfo {author} {\bibfnamefont {R.~K.}\
  \bibnamefont {Kawakami}}, \bibinfo {author} {\bibfnamefont {A.}~\bibnamefont
  {Bostwick}}, \bibinfo {author} {\bibfnamefont {E.}~\bibnamefont {Rotenberg}},
  \ and\ \bibinfo {author} {\bibfnamefont {C.}~\bibnamefont {Jozwiak}},\ }\href
  {\doibase 10.1021/acsnano.6b04914} {\bibfield  {journal} {\bibinfo  {journal}
  {ACS Nano}\ }\textbf {\bibinfo {volume} {10}},\ \bibinfo {pages} {10058}
  (\bibinfo {year} {2016})}\BibitemShut {NoStop}%
\bibitem [{\citenamefont {Wilson}\ \emph {et~al.}(2017)\citenamefont {Wilson},
  \citenamefont {Nguyen}, \citenamefont {Seyler}, \citenamefont {Rivera},
  \citenamefont {Marsden}, \citenamefont {Laker}, \citenamefont
  {Constantinescu}, \citenamefont {Kandyba}, \citenamefont {Barinov},
  \citenamefont {Hine}, \citenamefont {Xu},\ and\ \citenamefont
  {Cobden}}]{Wilson_SciAdv_2017}%
  \BibitemOpen
  \bibfield  {author} {\bibinfo {author} {\bibfnamefont {N.~R.}\ \bibnamefont
  {Wilson}}, \bibinfo {author} {\bibfnamefont {P.~V.}\ \bibnamefont {Nguyen}},
  \bibinfo {author} {\bibfnamefont {K.}~\bibnamefont {Seyler}}, \bibinfo
  {author} {\bibfnamefont {P.}~\bibnamefont {Rivera}}, \bibinfo {author}
  {\bibfnamefont {A.~J.}\ \bibnamefont {Marsden}}, \bibinfo {author}
  {\bibfnamefont {Z.~P.~L.}\ \bibnamefont {Laker}}, \bibinfo {author}
  {\bibfnamefont {G.~C.}\ \bibnamefont {Constantinescu}}, \bibinfo {author}
  {\bibfnamefont {V.}~\bibnamefont {Kandyba}}, \bibinfo {author} {\bibfnamefont
  {A.}~\bibnamefont {Barinov}}, \bibinfo {author} {\bibfnamefont {N.~D.~M.}\
  \bibnamefont {Hine}}, \bibinfo {author} {\bibfnamefont {X.}~\bibnamefont
  {Xu}}, \ and\ \bibinfo {author} {\bibfnamefont {D.~H.}\ \bibnamefont
  {Cobden}},\ }\href {\doibase 10.1126/sciadv.1601832} {\bibfield  {journal}
  {\bibinfo  {journal} {Sci. Adv.}\ }\textbf {\bibinfo {volume} {3}},\ \bibinfo
  {pages} {e1601832} (\bibinfo {year} {2017})}\BibitemShut {NoStop}%
\bibitem [{\citenamefont {Chen}\ \emph {et~al.}(1990)\citenamefont {Chen},
  \citenamefont {Fritze}, \citenamefont {Nurmikko}, \citenamefont {Ackley},
  \citenamefont {Colvard},\ and\ \citenamefont {Lee}}]{Chen_PRL_1990}%
  \BibitemOpen
  \bibfield  {author} {\bibinfo {author} {\bibfnamefont {W.}~\bibnamefont
  {Chen}}, \bibinfo {author} {\bibfnamefont {M.}~\bibnamefont {Fritze}},
  \bibinfo {author} {\bibfnamefont {A.~V.}\ \bibnamefont {Nurmikko}}, \bibinfo
  {author} {\bibfnamefont {D.}~\bibnamefont {Ackley}}, \bibinfo {author}
  {\bibfnamefont {C.}~\bibnamefont {Colvard}}, \ and\ \bibinfo {author}
  {\bibfnamefont {H.}~\bibnamefont {Lee}},\ }\href {\doibase
  10.1103/PhysRevLett.64.2434} {\bibfield  {journal} {\bibinfo  {journal}
  {Phys. Rev. Lett.}\ }\textbf {\bibinfo {volume} {64}},\ \bibinfo {pages}
  {2434} (\bibinfo {year} {1990})}\BibitemShut {NoStop}%
\bibitem [{\citenamefont {Scuri}\ \emph {et~al.}(2018)\citenamefont {Scuri},
  \citenamefont {Zhou}, \citenamefont {High}, \citenamefont {Wild},
  \citenamefont {Shu}, \citenamefont {De~Greve}, \citenamefont {Jauregui},
  \citenamefont {Taniguchi}, \citenamefont {Watanabe}, \citenamefont {Kim},
  \citenamefont {Lukin},\ and\ \citenamefont {Park}}]{Scuri_PRL_2018}%
  \BibitemOpen
  \bibfield  {author} {\bibinfo {author} {\bibfnamefont {G.}~\bibnamefont
  {Scuri}}, \bibinfo {author} {\bibfnamefont {Y.}~\bibnamefont {Zhou}},
  \bibinfo {author} {\bibfnamefont {A.~A.}\ \bibnamefont {High}}, \bibinfo
  {author} {\bibfnamefont {D.~S.}\ \bibnamefont {Wild}}, \bibinfo {author}
  {\bibfnamefont {C.}~\bibnamefont {Shu}}, \bibinfo {author} {\bibfnamefont
  {K.}~\bibnamefont {De~Greve}}, \bibinfo {author} {\bibfnamefont {L.~A.}\
  \bibnamefont {Jauregui}}, \bibinfo {author} {\bibfnamefont {T.}~\bibnamefont
  {Taniguchi}}, \bibinfo {author} {\bibfnamefont {K.}~\bibnamefont {Watanabe}},
  \bibinfo {author} {\bibfnamefont {P.}~\bibnamefont {Kim}}, \bibinfo {author}
  {\bibfnamefont {M.~D.}\ \bibnamefont {Lukin}}, \ and\ \bibinfo {author}
  {\bibfnamefont {H.}~\bibnamefont {Park}},\ }\href {\doibase
  10.1103/PhysRevLett.120.037402} {\bibfield  {journal} {\bibinfo  {journal}
  {Phys. Rev. Lett.}\ }\textbf {\bibinfo {volume} {120}},\ \bibinfo {pages}
  {037402} (\bibinfo {year} {2018})}\BibitemShut {NoStop}%
\end{thebibliography}
\end{document}

\title{Supplemental Material for \\ ``Shubnikov-de Haas oscillations in optical conductivity of monolayer MoSe$_2$''}

\author{\surname{T.~Smole\'nski}}\affiliation{\ETH}\affiliation{\UW}
\author{\surname{O. Cotlet}}\affiliation{\ETH}
\author{\surname{A. Popert}}\affiliation{\ETH}
\author{\surname{P. Back}}\affiliation{\ETH}
\author{\surname{Y. Shimazaki}}\affiliation{\ETH}
\author{\surname{P.~Kn\"uppel}}\affiliation{\ETH}
\author{\surname{N.~Dietler}}\affiliation{\ETH}
\author{\surname{T.~Taniguchi}}\affiliation{\NIMS}
\author{\surname{K.~Watanabe}}\affiliation{\NIMS}
\author{\surname{M.~Kroner}}\affiliation{\ETH}
\author{\surname{A.~Imamoglu}}\affiliation{\ETH}

\maketitle
\section{Sample and Experimental Setup}
\label{sec:sample_setup}

The device investigated in the main text consists of a MoSe$_2$ monolayer, which is contacted with a few-layer graphene (FLG) flake and fully encapsulated between two thin films of hexagonal boron-nitride ($h$-BN). These films are sandwiched between a pair of FLG flakes serving as top and back gates, out of which only the top gate was utilized to control the carrier density in the present study. All of the flakes were mechanically exfoliated from the bulk crystals (HQ Graphene MoSe$_2$, NIMS $h$-BN, and natural graphite) using wafer dicing tape (Ultron) and then subsequently transferred onto Si substrates covered by a 285~nm thick SiO$_2$ layer. At this state the quality and thicknesses of the flakes were verified by optical contrast measurements~\cite{Ni_NL_2007, Li_ACSNano_2013, Golla_APL_2013} and/or atomic force microscopy (AFM). Moreover, the thicknesses of top and bottom $h$-BN layers, yielding respectively $t_\mathrm{t}=(81\pm5)$~nm and $t_\mathrm{b}=(24\pm3)$~nm, were deliberately chosen based on transfer-matrix simulations of the device reflectance spectrum~\cite{Back_PRL_2018} in order to ensure that: (1) the excitonic resonance exhibits approximately Lorentzian lineshape, (2) the MoSe$_2$ monolayer is placed around a node of the optical field confined inside an effective cavity formed by the heterostucture, with the aim to prolong the radiative lifetime of the exciton~\cite{Scuri_PRL_2018}, and hence to reduce the linewidth of its optical transition. 

\begin{figure*}
\includegraphics{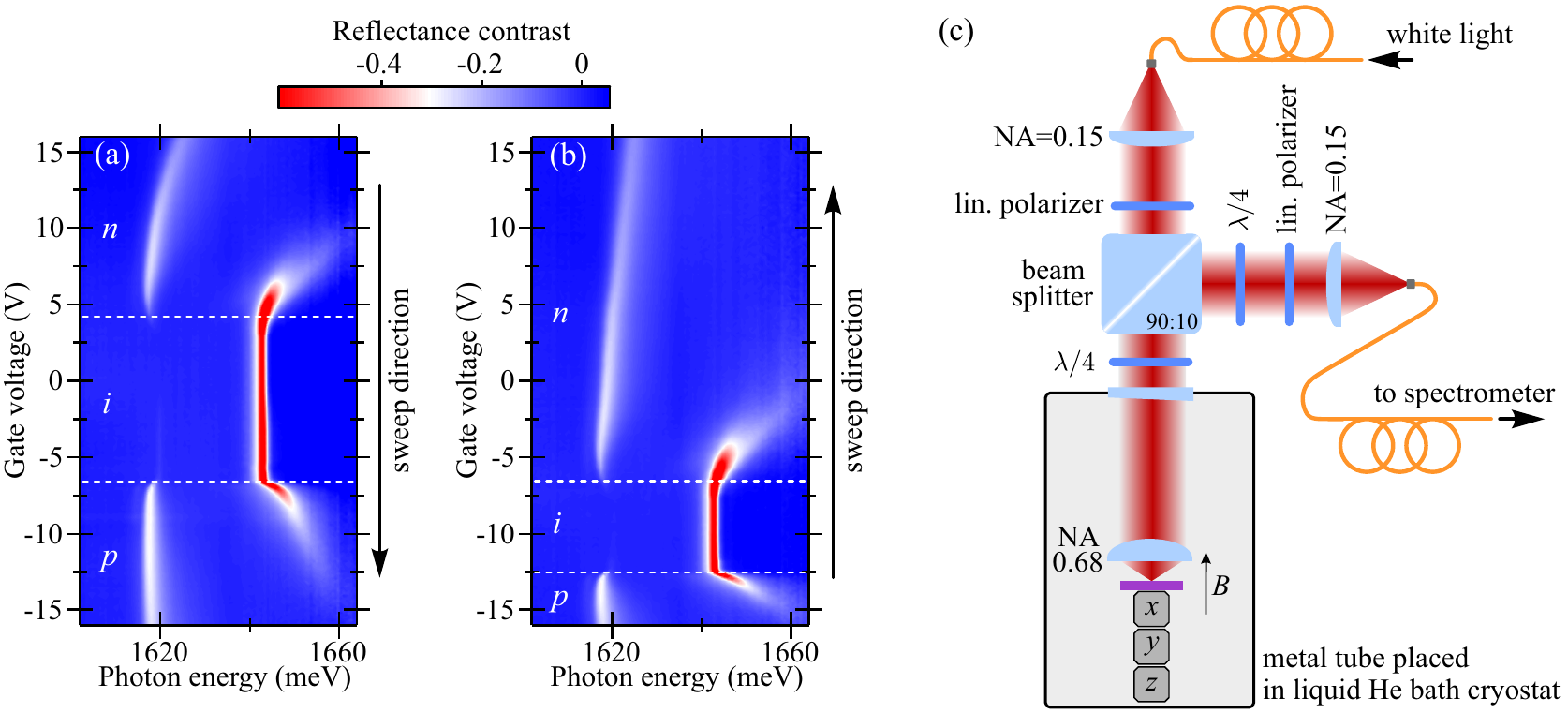}
\caption{(a,b) Color-scale maps presenting the zero-magnetic-field reflectance contrast spectra measured as a function of the gate voltage $V_g$ at the same spot on the sample, but with opposite gate sweep directions: (a) from +16~V to -16~V and (b) from -16~V to +16~V. The horizontal dashed lines mark the crossovers between neutral and $n$- or $p$-doped regimes (as indicated). (c)~Simplified schematic of the experimental setup. \label{fig:histeresis_and_exp_setup}}
\end{figure*}

The multilayer stack was assembled by means of a dry-transfer method~\cite{Zomer_APL_2014, Pizzocchero_NC_2016} involving the use of a glass slide holding a hemispherical polydimethylsiloxane (PDMS) stamp~\cite{Kim_NL_2016} covered with a thin polycarbonate (PC) layer to sequentially pick up the flakes. All stacking steps were performed inside the glove box in inert Ar atmosphere. The stacking was carried out in a high temperature of 120$^\circ$C, which, together with a slow stacking speed, allowed us to reduce the number of contamination pockets~\cite{Pizzocchero_NC_2016}. In this regard, the utilized hemispherical shape of the PDMS stamp was a helpful aid, as it facilitated the control over the size of the contact front between the PC and the substrate. The finished stack was released onto the SiO$_2$/Si substrate, and afterwards the residual PC film was removed from its surface by immersing in chloroform. No annealing was performed during the whole fabrication process. Finally, the FLG flakes were electrically contacted with metal electrodes prepared by standard electron-beam lithography and subsequent deposition of 105~nm~thick gold layer on top of a 5~nm~thick titanium sticking layer. The MoSe$_2$ monolayer was doped by applying a voltage $V_g$ between the FLG top gate and the FLG contact to the monolayer. The resulting carrier density was, however, found to be determined not only by the value of $V_g$, as the device displayed a pronounced hysteresis when $V_g$ was swept in a loop [see Figs~\ref{fig:histeresis_and_exp_setup}(a,b)], which presumably stems from a Schottky nature of the contacts. In order to avoid such a hysteretic behavior, in all of the experiments reported in the main text the gate voltage was always varied within a fixed range, at a constant rate, and in the same direction (i.e., from $V_g=16$~V to $-16$~V). In this way we were able to almost completely mitigate the impact of gate hysteresis on the measured quantities, as confirmed, e.g., by nearly perfect reproducibility of the $V_g$ values corresponding to the onset of filling the valence band with holes, which for all gate-voltage scans varied by less than 0.1~V.

Our magneto-optical experiments were carried out in a high-resolution, confocal microscope setup schematically depicted in Fig.~\ref{fig:histeresis_and_exp_setup}(c). The sample was mounted on \textit{x-y-z} piezo-electric stages inside a stainless steel tube filled with 20~mbar helium exchange gas at $T\approx4$~K. The tube was immersed in a liquid helium bath cryostat equipped with a superconducting coil generating a magnetic field of up to 16~T in Faraday geometry. A free-space optical access to the sample was provided by a wedged window on top of the tube. The reflectance measurements were performed with the use of a single-mode-fiber-coupled broadband light emitting diode (LED) with a center wavelength of 760~nm and a 3~dB bandwidth of 20~nm. The excitation light, after exiting the fiber, was collimated before it entered the tube, where it was focused on the sample surface by a high numerical aperture aspheric lens (NA~=~0.68). The excitation power was kept in the range of a few tens of~nW. The light reflected off the sample was collected by the same lens, separated from incident light by a beam splitter, and finally coupled to a single-mode fiber that guided it to 0.5~m spectrometer equipped with a liquid nitrogen-cooled charge-coupled-device (CCD), which was utilized to analyze the reflectance spectra. A set of polarization optics, including linear polarizers and quarter wave-plates, was incorporated in both excitation and detection paths in order to linearly polarize the incident beam as well as to detect the reflected light in $\sigma^+$ or $\sigma^-$ circular polarizations.

\section{Analysis of the reflectance spectra}

\begin{figure*}[b]
\includegraphics{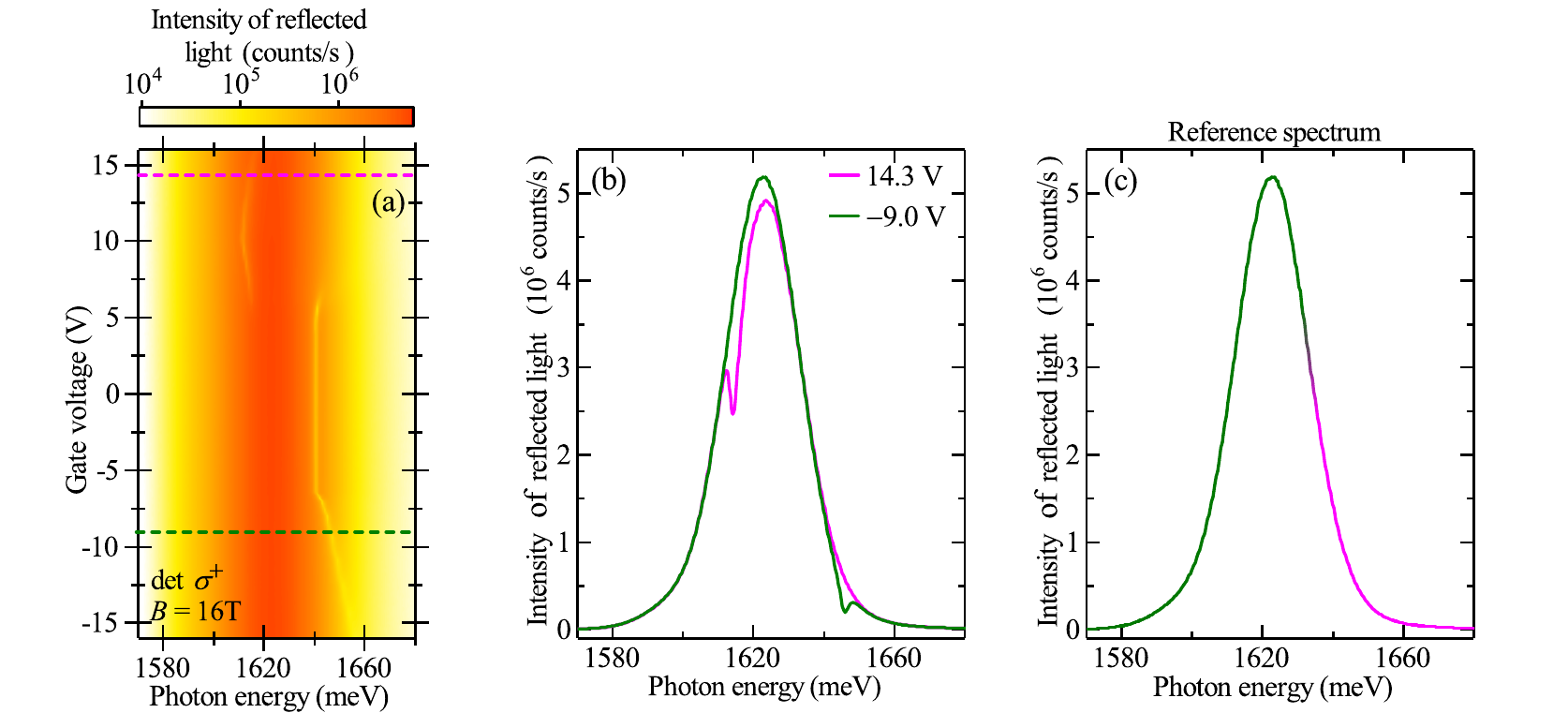}
\caption{Construction of the reference spectrum from reflectance measurements. (a) Logarithmic-color-scale map presenting example (unnormalized) $\sigma^+$-polarized gate-voltage-dependent spectra of the LED light reflected off the sample at $B=16$~T. (b) Line-cuts through the data in (a) for two different gate voltages, showing the spectra featuring only one resonance that originates either from the exciton (green curve) or attractive polaron (pink curve). (c) The reference spectrum constructed by combing low-energy (green color) and high-energy (pink color) resonance-free spectral regions from, respectively, the green and pink spectrum in (b).\label{fig:reference_spectrum}}
\end{figure*}

To obtain the reflectance contrast from the measured spectra of the LED light reflected off the sample it is necessary to determine the unperturbed reflection spectrum of the LED. In principle, such a reference spectrum may be obtained by moving the excitation spot off the MoSe$_2$ monolayer region on the sample surface. However, due to inevitable spatial inhomogeneity of the heterostructure, this procedure is often fraught with a systematic error. For this reason we follow a more reliable approach, in which the reference spectrum is constructed based on the reflectance spectra taken at a fixed spot, but for different gate voltages. More specifically, from a given circular-polarization-resolved gate-voltage scan we select two spectra in such a way that each of them exhibits only one resonance corresponding either to the exciton or attractive polaron, as shown in Figs~\ref{fig:reference_spectrum}(a,b) for an example data set taken at $B=16$~T. Owing to narrow linewidths and large energy splitting of these optical transitions, such a spectra feature partially overlapping, resonance-free spectral regions, which are combined together to obtain an unperturbed reference spectrum $R_0(E)$ [Fig.~\ref{fig:reference_spectrum}(c)]. On this basis we evaluate the reflectance contrast spectra $R_c(E)=[R(E)-R_0(E)]/R_0(E)$ from the reflectance spectra $R(E)$ measured at each gate voltage ($E$ represents the photon energy).

\begin{figure*}[t]
\includegraphics{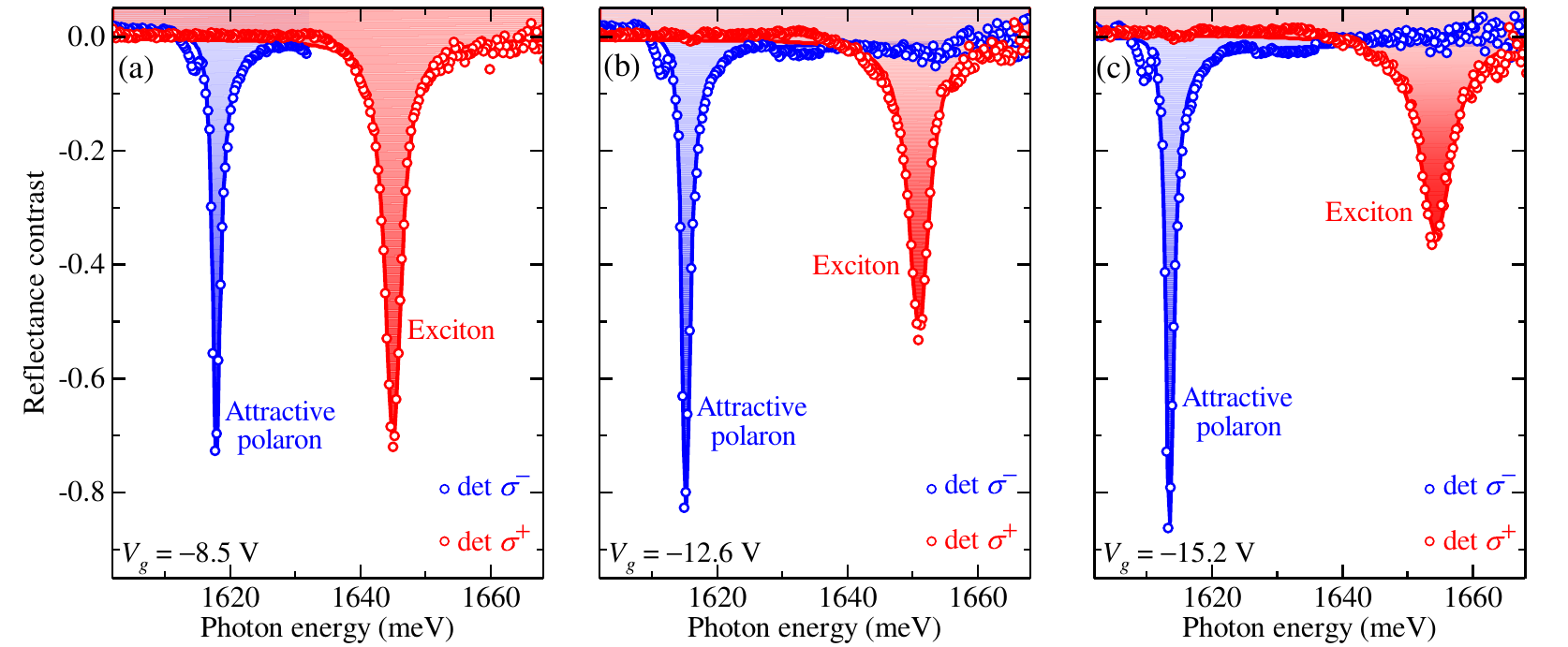}
\caption{(a-c) Reflectance contrast spectra measured at $B=16$~T in $\sigma^+$ (red) and $\sigma^-$ (blue) circular polarizations for three different gate voltages $V_g$ corresponding to hole-doped regime: $-8.5$~V (a), $-12.6$~V (b), and $-15.2$~V (c). Solid lines represent the fits of the exciton and attractive polaron resonances (visible, respectively, in $\sigma^+$ and $\sigma^-$ polarizations) with the formula described by Eq.~(\ref{eq:fitting_formula}).\label{fig:fit_X_AP_resonances}}
\end{figure*}

In order to determine the transition energy and the linewidth of the exciton and attractive polaron resonances in the hole-doped regime we first recall that the reflectance contrast measured in our experiments is not only determined by the MoSe$_2$ monolayer optical susceptibility $\chi(E)$. Instead, it can be effectively described as $\mathrm{Im}[e^{i\alpha(E)}\chi(E)]$, where $\alpha(E)$ stands for a wavelength-dependent phase-shift arising from the interferences of light reflected at different interfaces (e.g., \mbox{$h$-BN/SiO$_2$}) of the heterostructure~\cite{Back_PRL_2017}. A more comprehensive analysis of the lineshape can be carried out using the transfer matrix method; however, since the absolute strength of the susceptibility is not of interest in this work, we carry out the outlined, simpler approach. Due to a suitable choice of the thicknesses of top and bottom $h$-BN films (as described in Sec.~\ref{sec:sample_setup}), the invoked phase-shift is essentially close to $180^\circ$ within the spectral region of interest in case of our device. In particular, $\alpha\simeq180^\circ$ for the exciton resonance, as it features almost purely Lorentzian lineshape [see \mbox{Figs~\ref{fig:fit_X_AP_resonances}(a-c)]} that, except for forming a dip, exactly corresponds to the expected Lorentzian spectral profile of the imaginary part of the susceptibility. We remark here that, in general, the assumed Lorentzian lineshape is only valid in the absence of free carriers, whereas for non-zero hole densities the spectral profile of the susceptibility may be modified by the exciton-hole interactions~\cite{Sidler_NatPhys_2017, Back_PRL_2018}. However, we do not find any clear signatures of the exciton lineshape variation with the gate voltage. As seen in Figs~\ref{fig:fit_X_AP_resonances}(a-c), the attractive polaron resonance appearing in the opposite circular polarization, exhibits a slightly asymmetric lineshape. As revealed by our transfer-matrix simulations, this asymmetry originates from a combined effect of a reduced monolayer reflectivity, different wavelength and narrower linewidth of this resonance, and hence it can be accounted for by setting $\alpha\simeq169^\circ$ for the attractive polaron instead of $\alpha\simeq180^\circ$ as in the case of the exciton.

Based on the above considerations, we describe the reflectance contrast spectral profiles of both resonances with the following formula:
\begin{equation}
\label{eq:fitting_formula}
R_c(E)=A\cos(\alpha)\frac{\gamma/2}{(E-E_0)^2+\gamma^2/4}+A\sin(\alpha)\frac{E_0-E}{(E-E_0)^2+\gamma^2/4}+C,
\end{equation}
where $A$, $E_0$, and $\gamma$ represent, respectively, the amplitude, energy, and linewidth of a given resonance, while the parameter $C$ is introduced to capture any broad background contribution to the measured signal. Using this formula we are able to almost perfectly reproduce the experimental data for different carrier densities, as shown by the solid lines in Figs~\ref{fig:fit_X_AP_resonances}(a-c). From such fits we extract the energies and linewidths of the exciton/polaron resonances as a function of the gate voltage, which, in particular, are plotted in Figs~3(a,b) in the main text. To evaluate the derivatives of the energies $E$ with respect to the voltage $V_g$ [presented in Fig. 3(c)], the extracted data are binned in 80~mV intervals, and $|\mathrm{d}E/\mathrm{d}V_g|$ is subsequently computed as an absolute value of the difference quotient between neighboring data points. The voltages $V_g(\nu)$ corresponding to integer filling factors $\nu$ [shown in Figs 3(d) and 5] are in turn determined based on the oscillations of the exciton linewidth. To this end, the original data are first convolved with a Gaussian of standard deviation 30~mV in order to reduce the noise, and then the searched voltages $V_g(\nu)$ are determined as those corresponding to the data points with locally minimal linewidth. 

\section{Capacitive model for evaluation of the carrier density}

During our experiments, the hole density $p$ in the MoSe$_2$ monolayer was tuned by applying a gate voltage $V_g$ between the FLG top gate and the FLG contact to the monolayer. In order to establish a link between the voltage~$V_g$ and the density~$p$, we employ a simple capacitive model, in which the effective capacitance per unit area of the device is given by
\begin{equation}
C=(C_\mathrm{Q}^{-1}+C_\mathrm{geom}^{-1})^{-1},
\end{equation}
where $C_\mathrm{Q}=e^2D(E_F)$ denotes the quantum capacitance determined by the density $D(E_F)$ of electronic states (DOS) in the MoSe$_2$ at the Fermi level $E_F$, while $C_\mathrm{geom}$ stands for a geometric contribution, which, in the parallel-plate approximation, can be described~as
\begin{equation}
C_\mathrm{geom}=\frac{\epsilon_0\epsilon^\perp_{h\mathrm{-BN}}}{t_{h\mathrm{-BN}}}.
\label{eq:C_geom}
\end{equation}
Here $t_{h\mathrm{-BN}}=(81\pm5)$~nm represents the thickness of the top $h$-BN layer separating the top gate and the MoSe$_2$ monolayer, whereas $\epsilon^\perp_{h\mathrm{-BN}}$ denotes the perpendicular, static dielectric constant of $h$-BN, which is estimated as $(3.5\pm0.5)$ based on the values provided in several previous reports~\cite{Kim_ACSNano_2012,Laturia_2DMA_2018}. 

In this model, the quantum capacitance plays an important role only if $C_\mathrm{Q}\ll C_\mathrm{geom}$, that is when $D(E_F)$ is negligibly small (e.g., in the band gap). In such a case the carrier density remains constant, while the change of the gate voltage results solely in the shift of the Fermi level by $\mathrm{d}E_F/\mathrm{d}V_g=e$. Concurrently, in the opposite regime of large $D(E_F)$, which is realized when the Fermi level lies in the valence or conduction bands, the quantum capacitance $C_\mathrm{Q}\gg C_\mathrm{geom}$ can in turn be neglected, and the carrier density increases linearly with the gate voltage according to $\mathrm{d}p/\mathrm{d}V_g=-C_\mathrm{geom}/e$. We note here that when the band states break up into a series of Landau Levels (LLs) upon application of an external magnetic field $B$, the condition $C_\mathrm{Q}\gg C_\mathrm{geom}$ does not necessarily have to be fulfilled for the Fermi level lying in the gap between the LLs. However, for a realistic device, owing to inevitable disorder-induced LL broadening, the DOS in the inter-LL gap remains presumably large enough for the invoked condition to hold. But even in an ideal case of delta-like energy spectrum of Landau-quantized DOS, the deviations from a linear dependence of the carrier density on the gate voltage still turn out to be negligible. More specifically, in such a scenario the invoked dependence would exhibit a step-like character, featuring a periodic linear-increase regions separated by plateaus, each extending over the gate voltage range $\Delta V_\mathrm{inter}=E_c/e$, in which the Fermi level moves across the inter-LL cyclotron energy gap $E_c=\hbar eB/m^*$, where $m^*$ is the carrier effective mass. Crucially, this voltage range remains orders of magnitude narrower than the range of $\Delta V_\mathrm{LL}=\Delta p_\mathrm{LL} e/C_\mathrm{geom}$, in which the Fermi level is pinned to a LL that is gradually filled with holes, where $\Delta p_\mathrm{LL}=feB/h$ denotes the number of states in the LL. This can be readily seen by computing the ratio $\Delta V_\mathrm{inter}/\Delta V_\mathrm{LL}=2\pi\hbar^2fC_\mathrm{geom}/e^2m^*$, which for $f=1$ and $m^*\approx0.5m_0$~\cite{Liu_PRB_2013, Kormanyos_2DMat_2015} yields $2\cdot10^{-3}$, showing that nonlinearities in the dependence of $p$ on $V_g$ related to DOS oscillations can indeed be safely neglected.

In light of the above theoretical analysis, the change of the gate voltage $\Delta V_\mathrm{LL}$ required to fill a LL with holes should be independent of the filling factor~$\nu$. However, according to our experimental results, this prediction remains valid only for appropriately large $\nu$ [e.g., $\nu>3$ for the 16~T data depicted in Fig.~3(d) in the main text], whereas at low hole densities ($p\lesssim1\cdot10^{12}$~cm$^{-2}$) $\Delta V_\mathrm{LL}$ is clearly getting smaller with decreasing $\nu$. We attribute these deviations to the presence of a considerable Schottky barrier at the interface of the MoSe$_2$ monolayer and the FLG contact, which is not taken into account in the above-described capacitive model, but turns out to play an important role for the main device. In particular, such a barrier prevents the holes from flowing into the monolayer, making it necessary to apply lower (i.e., more negative) gate voltages in order to reach the hole-doped regime. This, for example, entails larger than expected separation between the onsets of filling the valence and conduction bands, which yields approximately~10~V when the voltage is being swept towards negative values [see Fig.~\ref{fig:histeresis_and_exp_setup}(a)]. Once the Schottky barrier is overcome (at $V_g\approx-6.5$~V), the holes are abruptly injected to the monolayer with the rate $|\mathrm{d}p/\mathrm{d}V_g|$ much larger than the one of $C_\mathrm{geom}/e$ predicted by the capacitive model, resulting in a decreased value of $\Delta V_\mathrm{LL}$ as well as relatively rapid blue (red) shift of the exciton (attractive polaron) resonances in the reflectance spectra. As the gate voltage is further ramped down towards more negative values, the Schottky effects become less and less important, which in turn leads to a gradual reduction of the hole injection rate. Finally, at $V_g\approx-8$~V, we enter a linear-response regime, in which the injection rate saturates at a constant value, and further changes $\Delta p$ of the hole density can be described within the frame of the capacitive model as $-C_\mathrm{geom}\Delta V_g/e$. This is fully confirmed by a perfect agreement between the gate voltages $V_g(\nu,B)$ corresponding to integer $\nu$ that were extracted from the data in the invoked range and a set of linear dependencies of the form $V_g(\nu,B)=V_0-\delta\nu B$, as shown in Fig.~5 in the main text. Based on the value of the parameter $\delta=\Delta V_\mathrm{LL}/B=fe^2/C_\mathrm{geom}h$ controlling the slopes of these dependencies, we extract the LL degeneracy $f=1.0\pm0.2$, as stated in the main text. Moreover, by extrapolating these dependencies to zero magnetic field, we determine the gate voltage $V_0=-2.57$~V that would correspond to the onset of filling the valence band with holes in the absence of the Schottky effects. This allows us to obtain the absolute value of the hole density for $V_g\lesssim-8$~V [marked on the right vertical axis in Fig.~5 in the main text] using the following relation:
\begin{equation}
p(V_g\lesssim-8\ \mathrm{V})=\frac{C_\mathrm{geom}}{e}(V_0-V_g)\approx(2.39\pm0.49)\cdot10^{11}\ \mathrm{\frac{cm^{-2}}{V}}\cdot(V_0-V_g).
\end{equation}

\section{Reproducibility of the results on a different device}
\label{sec:second_sample}

\begin{figure*}[b]
\includegraphics{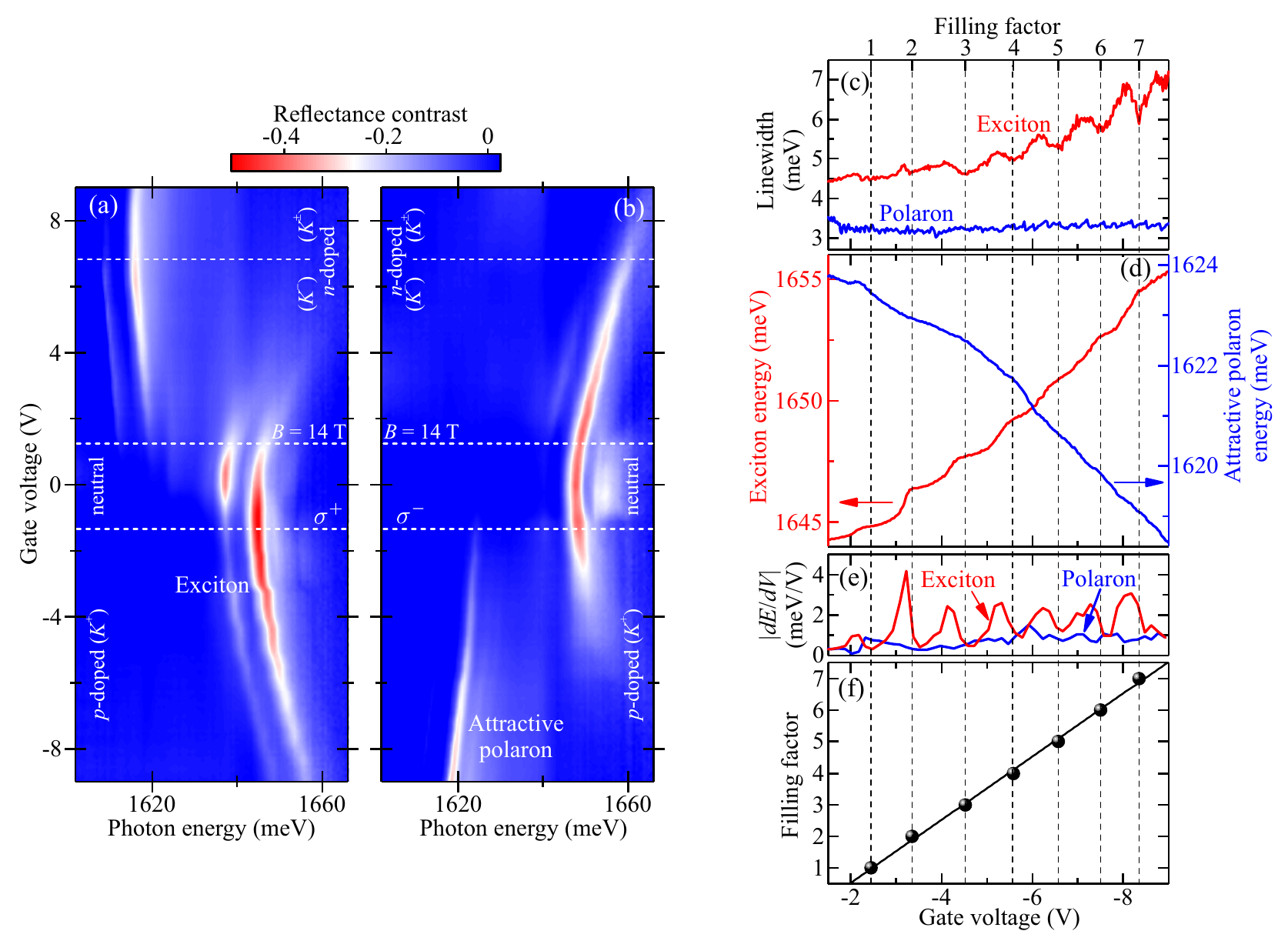}
\caption{(a,b) Color-scale maps presenting the reflectance contrast spectra measured for the second device as a function of the back gate voltage $V_g$ at $B=14$~T in $\sigma^+$ (a) and $\sigma^-$ (b) circular polarization. Dashed lines mark the crossovers between different doping regimes (as indicated). (c-e) Gate-voltage dependencies of the exciton/attractive polaron linewidths (c), energies (d), and derivatives of the energies with respect to $V_g$ (e). The former two quantities were extracted from the data in (a,b) by fitting the main (i.e., most intense) resonances with the formula given by Eq.~(\ref{eq:fitting_formula}) for $C=0$ and $\alpha=158^\circ$. To avoid spurious effects related to the presence of lower-intense side peaks, the fitting was carried out within a 10~meV (6~meV) wide spectral window around the exciton (attractive polaron) resonance. Prior determination of the derivatives, the data were binned in 150~mV intervals in order to reduce the noise. (f)~The gate voltages corresponding to integer filling factors obtained based on the positions of the local minima of the exciton linewidth (for determining these positions, the original data were first convolved with a Gaussian of standard deviation 60~mV). The solid line represents the linear fit to the data.\label{fig:data_second_device}}
\end{figure*}

As stated in the main text, our experiments have been repeated on a second van der Waals heterostructure yielding consistent results. This heterostructure was fabricated using very similar technique to that utilized for preparation of the device studied in the main text (see Sec.~\ref{sec:sample_setup}). It consisted of a charge-tunable MoSe$_2$ monolayer, which was electrically contacted with a FLG flake, encapsulated between two layers of $h$-BN, and finally sandwiched between another two FLG flakes serving as top and back gates, out of which only the back gate was utilized to tune the carrier density in the present experiments. Figs~\ref{fig:data_second_device}(a,b) show circular-polarization-resolved reflectance contrast spectra measured for this device as a function of the back gate voltage under a magnetic field of $B=14$~T. The data are of clearly lower quality than those obtained for the main device. In particular, the exciton/polaron resonances are significantly broader and accompanied by several lower-intense side peaks, which most probably appear due to the contributions of distinct, simultaneously excited sub-micrometre sized regions lying within the excitation spot. Nonetheless, both the main and side exciton resonances demonstrate qualitatively similar behavior to that seen for the main sample, featuring a pronounced Shubnikov-de Haas-like oscillations on the hole-doped side (i.e., in $\sigma^+$ polarization for $V_g\lesssim-1.5$~V) that are certainly much fainter at the electron doping (i.e., in $\sigma^-$ polarization for $V_g\gtrsim1.5$~V).

In order to quantitatively compare the behavior of the exciton/polaron resonances for the two devices, the spectral characteristics of the main (i.e., most intense) resonances extracted for the second sample on the hole-doped side are plotted in~Figs~\ref{fig:data_second_device}(c-f) in the same format as in Figs~3(a-d) from the main text. As expected from the above analysis, the behavior is found to be indeed very similar, confirming the robustness of our conclusions drawn in the main text. In particular, the polaron linewidth remains almost constant with the gate voltage $V_g$, whereas the exciton linewidth is clearly increasing for more negative $V_g$. Additionally, it displays aforementioned, periodic oscillations related to sequential filling of the LLs, which are correlated with the oscillations of the exciton energy. However, in contrast to the main device, the signatures of such oscillations are not particularly evident in the gate-voltage dependence of the attractive polaron energy, which most probably stems from larger broadening of the resonances in the present device. 

Most importantly, the gate voltages $V_g(\nu)$ corresponding to integer filling factors $\nu$ (determined based on the positions of the local minima of the exciton linewidth) are almost equidistant, as revealed by a linear decrease of $V_g(\nu)$ with $\nu$ [see~\ref{fig:data_second_device}(f)]. This finding shows that the hole density remains proportional to $V_g$ in the whole experimentally accessible voltage range on the hole-doped side. This, in turn, indicates that the Schottky effects at the MoSe$_2$/FLG interface play much less important role in case of the second sample, which is further corroborated by much lower hysteresis observed for this device when sweeping $V_g$ in a loop. By fitting the values $V_g(\nu)$ with a linear curve, we obtain the voltage change $\Delta V_\mathrm{LL}=|V_g(\nu+1) - V_g(\nu)|=fe^2B/C_\mathrm{geom}h\approx1.00$~V needed to fill a LL. On this basis, as well as based on the value of a geometric capacitance $C_\mathrm{geom}$ evaluated using Eq.~(\ref{eq:C_geom}) for $t_{h\mathrm{-BN}}=(50\pm5)$~nm revealed by the AFM measurements of the bottom $h$-BN flake, we finally extract the LL degeneracy for the second device $f=1.1\pm0.3$, which is found to be equal to 1 within the experimental uncertainty, similarly to the case of the main device.

\section{Signatures of Landau levels filling in the optical spectra at electron doping}

\begin{figure*}[t]
\includegraphics{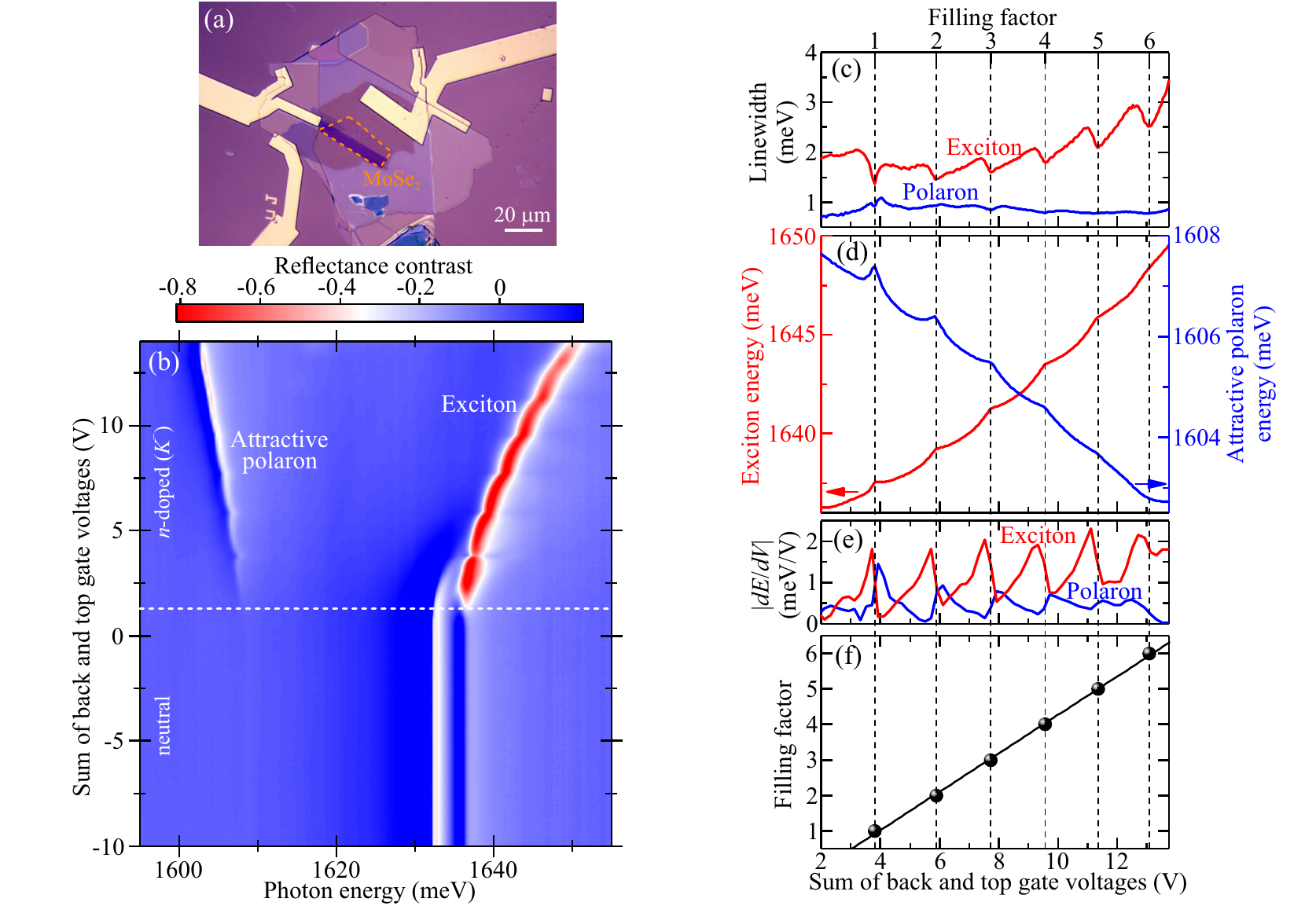}
\caption{(a) Optical micrograph of the third studied heterostructure exhibiting signatures of LL filling in the optical spectra on the electron-doped side. This device consisted of an dual-gated, $h$-BN-encapsulated MoSe$_2$ monolayer, whose boundaries are marked with a dashed line. (b) Color-scale map presenting the reflectance contrast spectra measured for this device as a function of the sum $V_g$ of back and top gate voltages at $B=16$~T in a linear polarization of detection. Dashed line marks the crossover between the neutral and valley-polarized $n$-doped regime, in which the electrons occupy only the states in the $K^-$ valley. (c-e) Gate-voltage dependencies of the exciton/attractive polaron linewidths (c), energies (d), and derivatives of the energies with respect to $V_g$ (e) determined at the electron-doped regime. The former two quantities were extracted from the data in (b) by fitting the exciton and attractive polaron resonances with the formula given by Eq.~(\ref{eq:fitting_formula}) for $\alpha=153^\circ$ and $\alpha=101^\circ$, respectively. To avoid the influence of other transitions in the spectra, the fitting was carried out within a 16~meV (20~meV) wide spectral window comprising the exciton (attractive polaron) resonance. Prior determination of the derivatives, the data were binned in 200~mV intervals in order to reduce the noise. (f)~The values of the sum of top and back gate voltages corresponding to integer filling factors obtained based on the positions of the local minima of the exciton linewidth (for determining these positions, the original data were first convolved with a Gaussian of standard deviation 50~mV). The solid line represents the linear fit to the data.\label{fig:data_third_device_electron_doping}}
\end{figure*}

One of the most surprising results of our analysis of the magneto-optical spectra acquired for each of two investigated devices (either the main one or that studied in Sec.~\ref{sec:second_sample}) is the fact that the signatures of LL filling were found to be prominent only on the hole-doped side, whereas at the electron doping they were much weaker, if present at all. As stated in the main text, we attribute this asymmetry to be at least partially due to a difference in effective masses of both carriers, with the electron mass being noticeably larger, as suggested by the results of recent transport experiments on Mo-based TMD monolayers~\cite{Larentis_PRB_2018,Pisoni_arXiv_2018}. This hypothesis would imply the electron LLs to feature lower cyclotron energy spacing, resulting in lower amplitude of the electron DOS modulation in the presence of inevitable disorder-induced LL broadening. Following this argument, one may anticipate analogous Shubnikov-de Haas-like oscillations to emerge also in the optical spectra on the electron-doped side for a better-quality device exhibiting lower LL broadening. We find that this is indeed the case, as revealed by the data obtained for the third device, whose optical micrograph is depicted in Fig.~\ref{fig:data_third_device_electron_doping}(a). This sample was fabricated in a similar way to the other two devices, following the procedure outlined in Sec.~\ref{sec:sample_setup}. It consisted of a charge-tunable MoSe$_2$ monolayer that was electrically contacted with two separate FLG flakes, encapsulated between two $h$-BN films, and finally sandwiched between top and bottom FLG gates, which were both utilized to control the carrier density in the monolayer. Due to very similar thicknesses $t=(80\pm10)$~nm of the two $h$-BN layers (extracted based on the optical contrast measurements), the geometrical capacitances $C_\mathrm{geom}$ [given by Eq.~(\ref{eq:C_geom})] between each gate and the MoSe$_2$ were almost identical, implying the electron density to be determined by the sum $V_g=V_\mathrm{top}+V_\mathrm{bottom}$ of the top and back gate voltages. 

Fig.~\ref{fig:data_third_device_electron_doping}(b) shows linearly-polarized reflectance contrast spectra measured at $B=16$~T on a particular spot on this device as a function of $V_g$ in the voltage range corresponding to either neutral or valley-polarized $n$-doped regime (in which the electrons occupy exclusively the states in $K^-$ valley). As seen, both the exciton and exciton-polaron resonances are remarkably sharp, featuring roughly two times narrower linewidth as compared to the corresponding resonances for the main device. It is most probably the reason for which the optical transitions observed for the present device display prominent Shubnikov-de Haas-like oscillations related to the filling of the electron LLs, which remained almost unresolved for previously investigated samples. Importantly, the character of these oscillations turns out to be almost identical to that established for the hole LLs in the main text, as proven by Figs~\ref{fig:data_third_device_electron_doping}(c-f) presenting the spectral characteristics of the exciton/polaron resonances for the present device in the same format as in Figs \mbox{3(a-d)} for the case of the hole doping of the main device. First, the linewidth of the exciton transition undergoes familiar, periodic oscillations, whose sharp minima are perfectly correlated with cusp-like changes of the slopes of the exciton and attractive polaron energy dependencies on $V_g$. Second, these cusp-like features become clearly weaker at larger electron densities for the polaron. Finally, the gate voltages $V_g(\nu)$ associated with integer $\nu$ (i.e., with the minima of the exciton linewidth) are almost equidistant, with the voltage change needed to fill a LL $\Delta V_\mathrm{LL}=|V_g(\nu+1)-V_g(\nu)|=fe^2B/C_\mathrm{geom}h\approx1.85$~V corresponding to a LL degeneracy $f=1.2\pm0.3$ that is equal to 1 within the experimental uncertainty, exactly as in the case of the hole LL. Altogether, the above-demonstrated similarity between the optical signatures of the electron and hole LLs clearly indicates that the interaction-related mechanisms, invoked in the main text to explain the hole-doped case, are also responsible for LL-induced oscillations at electron doping, which generalizes the conclusions drawn in our work to the case of both types of carriers.

\section{Theoretical model}
In this section we develop a theoretical model of the exciton in a TMD monolayer hosting a dilute hole system subjected to a strong magnetic field. Similarly to the experiment, we focus only on the spin-valley-polarized regime, in which the itinerant holes occupy only the states in the $K^+$ valley, and consider optical excitations in the same valley. In such a case, the exciton will experience two main effects due to the presence of the hole system: (1) its energy will be modified owing to phase-space filling; (2) it will acquire a finite correlation energy and lifetime owing to the interaction with the itinerant holes.

\subsection{Single particle Hamiltonian}
In the presence of a magnetic field, it is useful to use the basis of single particle eigenstates. In the symmetric gauge $\boldsymbol{A}=B(-y/2,x/2,0)$, which is employed here, these states can be labelled with two quantum numbers $\ket{nl}$, where $n\geq 0$ denotes the LL index, while $l\geq0$ is the canonical angular momentum of the state (with $l-n$ corresponding to the actual angular momentum). The single particle states $\ket{nl}$ can be generated using the ladder operators: 
\begin{eqnarray}
\ket{nl}=\frac{\left(a^\dagger\right)^n \left(b^\dagger\right)^l}{\sqrt{n! l!}}\ket{0 0},
\end{eqnarray}
starting from the ground state $\braket{\boldsymbol{r}}{00}=\frac{1}{\sqrt{2 \pi }}e^{-z z^*/4 }$, where $\boldsymbol{r}$ denotes the position of the particle. Notice that throughout this section we will use the complex representation of the position $z=x+i y$, and work in units of $\ell_B=\sqrt{\hbar/eB}=1$ and $\hbar=1$, where $\ell_B$ stands for the magnetic length, $e$ is the electron charge, while $\hbar$ the reduced Planck constant. In the above expression, $a^\dagger = (\Pi_x + i \Pi_y)/\sqrt{2}$ and $b^\dagger = (\Gamma_x - i \Gamma_y)/\sqrt{2}$ , where $\boldsymbol{\Pi}= - i \nabla_{\boldsymbol{r}} + e \boldsymbol{A} (\boldsymbol{r}) $ denotes the kinetic momentum, while $\boldsymbol{\Gamma}= - i  \nabla_{\boldsymbol{r}} - e \boldsymbol{A} (\boldsymbol{r})$ denotes the magnetic momentum. 

The single-particle kinetic energy Hamiltonian for a monolayer in magnetic field can be obtained using the Peierls substitution~\cite{Felix_PRB_2013,Tianyi_PRB_2013}. In case of the analyzed $K^+$ valley, this Hamiltonian can be described as: 
\begin{eqnarray}
H_0 = \sum_{n=0}^\infty \sum_{l=0}^{N_\phi-1} c^\dagger_{cnl} c_{cnl} \left[ \left(n +1\right) \omega_c +\Delta_g \right] + \sum_{n=0}^\infty \sum_{l=0}^{N_\phi-1}   c^\dagger_{vnl} c_{vnl} \left(-n \omega_c \right),
\end{eqnarray}
where $c^\dagger_{cnm}$ ($c^\dagger_{vnm}$) denotes the creation operator of an electron in the conduction (valence) band in the state $\ket{nm}$, $\Delta_g$ represents the semiconductor band gap, $\omega_c = eB/m^*$ stands for the cyclotron frequency (with $m^*$ representing the carrier effective mass, which assumed to be equal to 0.5$m_0$ for both valence and conductions bands), and $N_\phi = A/(2 \pi \ell_B^2)$ denotes the Landau level degeneracy with $A$ representing the quantization area. Notice that due to the chirality of the $K^+$ valley, the Landau levels in this valley are shifted up in energy by $\omega_c/2$ (while those in $K^{-}$ valley are shifted by the same energy in the opposite direction). We remark that we neglect here a small corrections to the wavefunctions/energies arising due to the Dirac nature of the material. This assumption is equivalent to neglecting deviations from the parabolic dispersion of the electrons in the absence of the magnetic field~\cite{Felix_PRB_2013,Tianyi_PRB_2013}. 

It is useful to introduce the electron and hole operators, such that: 
\begin{eqnarray}
& &e^\dagger_{nl} = c^\dagger_{c nl},\\
& &h^\dagger_{nl} = c_{v nl},
\end{eqnarray}
which allow us to write down the electron-hole kinetic energy Hamiltonian: 
\begin{eqnarray}
H_0 = \sum_{n=0}^\infty \sum_{l=0}^{N_\phi-1} e^\dagger_{nl} e_{nl} \left[ \left(n +1\right) \omega_c +\Delta_g \right] + \sum_{n=0}^\infty \sum_{l=0}^{N_\phi-1}   h^\dagger_{nl} h_{nl} n \omega_c.
\end{eqnarray}

\subsection{Exciton energy in a truncated Hilbert space}
In the absence of the electron, the holes will reside in a many-body ground state $\ket{\psi^h_{g}}$. It is beyond the scope of this Letter to accurately calculate this state. Instead, in the following simplified analysis we will assume that this state is an eigenstate of the angular momentum operator, and that it does not mix different LLs, i.e., $\bra{\psi^h_{g}} h^\dagger_{n'l'} h_{nl} \ket{\psi^h_{g}}=\nu_n \delta_{n'n}\delta_{l'l}$, where $\nu_n$ represents the filling of the $n$th hole LL. We further assume that $\ket{\psi^h_{g}}$ is a Gaussian state, which allows us to calculate any correlation functions using Wick's theorem. We expect this assumption to be justified due to the significant strength of Coulomb interactions in TMD monolayers. 

We are now going to calculate the energy of the exciton in the presence of the hole system. To this end, apart from the kinetic energy Hamiltonian $H_0$, we need also to take into account electron-hole Coulomb attraction as well as hole-hole Coulomb repulsion, which can be described using the following Hamiltonians:
\begin{eqnarray}
H_{eh}&=&-\sum_{q\sigma'\sigma\tau'\tau}\frac{V_{q}}{A} \bra{\sigma'}e^{-i\boldsymbol{q}\boldsymbol{r}_e}\ket{\sigma}\bra{\tau}e^{i\boldsymbol{q}\boldsymbol{r}_h}\ket{\tau'}e_{\sigma'}^\dagger h_{\tau'}^\dagger h_{\tau}e_{\sigma},\\
H_{hh}&=&\frac{1}{2}\sum_{q\tau_2'\tau_2\tau_1'\tau_1}\frac{V_{q}}{A} \bra{\tau_1}e^{-i\boldsymbol{q}\boldsymbol{r}_{h_1}}\ket{\tau_1'}\bra{\tau_2}e^{i\boldsymbol{q}\boldsymbol{r}_{h_2}}\ket{\tau_2'}h_{\tau_1'}^\dagger h_{\tau_2'}^\dagger h_{\tau_2} h_{\tau_1},
\end{eqnarray}
where $\boldsymbol{r}_e$ ($\boldsymbol{r}_h$) denotes the electron (hole) position, while $\boldsymbol{q}$ represents scattered momentum (in units of $1/\ell_B$), which will be further expressed in a complex representation $q=q_x+iq_y$. In the above formulas, $V_q$ denotes the Fourier transform of the Coulomb interaction, which in a two-dimensional system can be approximated as: 
\begin{eqnarray}
V_q = \frac{e^2}{4 \pi \epsilon_0 \frac{\epsilon_t+\epsilon_b}{2} \ell_B}  \frac{2 \pi }{|q| (1 + |q| \rho_0/\ell_B)},
\end{eqnarray}
where $\epsilon_t=\epsilon_b\approx3.5$ denotes the dielectric constant of the $h$-BN films on top and on the bottom of the TMD monolayer. The term of $\rho_0=4\pi \chi_{2D}/(\epsilon_t+\epsilon_b)$ accounts for the stronger screening inside the monolayer, where $\chi_{2D}$ represents the 2D polarizability of the planar material, which for MoSe$_2$ yields $\chi_{2D}=0.823$~nm~\cite{Berkelbach_PRB_2013}.

In order to determine the exciton ground state, we attempt to diagonalize full Hamiltonian $H=H_0+H_{eh}+H_{hh}$ in a truncated Hilbert space of a single electron-hole excitation, which is spanned by the states of the form $h^{\dagger}_{n'l'} e^\dagger_{nl} \ket{\psi^h_{g}}$. Owing to a significant size of this Hilbert space, a direct numerical calculation of the exciton binding energy is bound to fail. This can be readily seen by realizing that, due to strong Coulomb interactions, the exciton ground state is a superposition of many single-particle electron-hole excitations between electron and hole LLs of indices ranging up to $\sim(\ell_B/a_B^*)^2$, which yields about 40 for $B=16$~T and the exciton Bohr-radius of $a_B^*\approx1$~nm~\cite{Stier_NatCommun_2016}. Bearing in mind that we also need to include a similar number of angular momentum states in each LL, this finally leaves us with roughly $40^4$ states that would be required to properly calculate the exciton binding.

Fortunately, the above-introduced problem can be greatly simplified by taking advantage of the fact that we can eliminate the angular degrees of freedom due to quenching of the kinetic energy. This can be most conveniently done by introducing the density-like operators:
\begin{eqnarray}
\rho_e^{n'n}(k)&=& \sum_{l'l} e^{-|k|^2/2} G_k^{l'l} e^\dagger_{n'l'} e_{nl},\\
\rho_h^{n'n}(k)&=& \sum_{l'l} e^{-|k|^2/2}G_k^{ll'} h^\dagger_{n'l'} h_{nl} ,\label{eq:rho_h}\\
\rho_x^{n'n}(k)&=& \sum_{l'l} e^{-|k|^2/2}G_k^{l'l} e^\dagger_{n'l'} h^\dagger_{nl},
\end{eqnarray}
where $G$ functions are defined as~\cite{Efimkin_PRB_2018}: 
\begin{equation}
G^{n'n}_k=\left\{
\begin{array}{cc}
\sqrt{\frac{n!}{n'!}}\left(-\frac{k}{\sqrt{2}} \right)^{n'-n} L^{n'-n}_{n} \left(\frac{|k|^2}{2}\right) & ,\ \ \ \ n'\ge n,\\
\sqrt{\frac{n'!}{n!}}\left(\frac{k^*}{\sqrt{2}} \right)^{n-n'} L^{n-n'}_{n'} \left(\frac{|k|^2}{2}\right) & ,\ \ \ \ n' < n,
\end{array} \right.
\end{equation}
where $L_a^b$ represents the generalized Laguerre polynomials. These functions satisfy the following relations: 
\begin{eqnarray}
\label{i1}& & \left(G^{nm}_{k} \right) ^* = G^{mn}_{-k}, \\
\label{i3}& & \sum_{l=0}^\infty G^{m'l}_{k_1} G^{lm}_{k_2} = e^{-k_1^* k_2 /2} G^{m'm}_{k_1+k_2},\\
\label{i4}& & \sum_{m=0}^\infty G^{mm}_{k} = N_\phi \delta_{ k,0},\\
\label{i0}& & e^{-|k|^2/2}G_{k^*}^{n'n}G_{k}^{l'l}=\bra{n'l'}e^{-i\boldsymbol{k}\boldsymbol{r}}\ket{nl}.
\end{eqnarray}
More comprehensive description of the details of these functions is given in Refs~\cite{Efimkin_PRB_2018, MacDonald_arXiv_1994}. Once again we employ a complex notation such that $k=k_x+i k_y$. Using the above formulas, one can easily verify that the density operators obey the following algebra
\begin{eqnarray}
\ [\rho_x^{n'm'}  (k)^\dagger,\rho^{nm}_x(q)] &=& N_\phi \delta_{k,q} e^{-|k|^2/2} \delta_{n'n} \delta_{m'm}- \delta_{m'm} e^{-q k^*/2}\rho_e^{nn'}(q-k) -   \delta_{n'n} e^{-q^*k/2}\rho_h^{mm'}(q-k), \\ 
\ [\rho_e^{n'm'} (q),\rho^{nm}_x(k)] &=& \delta_{m'n} e^{q k^*/2} \rho_x^{n'm}(k+q),\\
\ [\rho_h^{n'm'} (q),\rho^{nm}_x(k)] &=& \delta_{m'm} e^{q^* k/2} \rho_x^{nn'}(k+q),\\
\ [\rho_x^{n'm'}(k)^\dagger, \rho_{e}^{nm}(q)]&=& \delta_{n'n} e^{-q^* k/2} \rho_x^{mm'} (k-q)^\dagger,\\
\ [\rho_x^{n'm'}(k)^\dagger, \rho_{h}^{nm}(q)]&=& \delta_{m'n} e^{-q k^*/2} \rho_x^{n'm} (k-q)^\dagger.
\end{eqnarray}
With the use of the above-defined density operators, we can re-express the Hamiltonian $H$: 
\begin{eqnarray}
H&=&H_0+H_{eh}+H_{hh},\\
H_0&=&\sum_{n} \omega_c \left[n  \rho^{nn}_h(0) +(n + 1 + \Delta_g/\omega_c)\rho_e^{nn}(0)  \right],\\
H_{eh}&=&-\sum_{qn'nm'm} \frac{V_q}{A} G^{n'n}_{q^*} G^{mm'}_{-q^*} \rho_e^{n'n}(q)\rho_h^{m'm}(-q),\\
H_{hh}&=&\frac{1}{2}\sum_{qn'nm'm} \frac{V_q}{A}  G^{nn'}_{q^*} G^{mm'}_{-q^*} \rho_h^{n'n}(q)\rho_h^{m'm}(-q) - \frac{1}{2}\sum_{qn} \frac{V_q}{A} \rho_h^{nn}(0).
\end{eqnarray}
We now make the exciton Ansatz: 
\begin{eqnarray}
x^\dagger_k \ket{\psi^h_{g}} \equiv \ket{k} = \sum_{nm} \phi^{nm}_k \frac{\rho_{x}^{nm}(k)}{\sqrt{\beta^{nm}_k}} \ket{\psi^h_{g}},
\label{eq_x_k}
\end{eqnarray}
where the summation runs over $n$ corresponding to (partially) empty hole LLs, while the introduced $\beta^{nm}_k$ is defined as:
\begin{eqnarray}
\beta^{nm}_k=\bra{\psi^h_{g}}\rho_x^{nm}(k)^\dagger \rho_x^{nm}(k) \ket{\psi^h_{g}} = N_\phi (1 - \nu_m) e^{-|k|^2/2}.
\end{eqnarray}
Notice that the normalization condition implies: 
\begin{eqnarray}
\braket{k'}{k}=\delta_{k'k}=\delta_{k'k}\sum_{nm} \left| \phi^{nm}_k \right|^2.
\end{eqnarray}
To obtain the ground-state exciton energy, we minimize $\bra{\psi^h_{g}} x_k (H - E) x^\dagger_k \ket{\psi^h_{g}}$, which yields the following expressions: 
\begin{eqnarray}
\sum_{n'm'} H^{nm}_{n'm'}(k) \phi_k^{n'm'}  =E \phi^{nm}_k \label{eq20},
\end{eqnarray}
where we introduced: 
\begin{eqnarray}
H^{n'm'}_{nm}(k)=H^{nm}_{n'm'}(k)^* &=& \frac{\bra{\psi^h_{g}} \rho_x^{n'm'} (k)^\dagger H \rho_{x}^{nm}(k)  \ket{\psi^h_{g}}}{\sqrt{\beta^{n'm'}_{k} \beta^{nm}_k}}.
\end{eqnarray}
The above expectation value can be expressed in terms of the following commutators: 
\begin{eqnarray}
\bra{\psi^h_{g}}  \rho_x^{n'm'} (k)^\dagger H  \rho_{x}^{nm}(k)  \ket{\psi^h_{g}} = \bra{\psi^h_{g}} \rho_x^{n'm'} (k)^\dagger [H,  \rho_{x}^{nm}(k) ] \ket{\psi^h_{g}} = \bra{\psi^h_{g}}  [\rho_x^{n'm'} (k)^\dagger, [H,  \rho_{x}^{nm}(k) ] ]  \ket{\psi^h_{g}},
\end{eqnarray}
where in the second step we assumed that $H\ket{\psi^h_{g}}=0$ (i.e., that the energy is calculated with respect to its value in the non-interacting case), while in the last step we took advantage of the fact that $\rho_x^{nm}(k)^\dagger \ket{\psi^h_{g}}=0$. We now calculate the expectation values of these commutators: 
\begin{eqnarray}
\frac{\bra{\psi^h_{g}}  [\rho_x^{n'm'} (k)^\dagger, [H_0,  \rho_{x}^{nm}(k) ] ]  \ket{\psi^h_{g}} }{\sqrt{\beta^{n'm'}_{k} \beta^{nm}_k}} &=&  \omega_c \left(n+m+1+\Delta_g/\omega_c\right) \delta_{n'n}\delta_{m'm}, \\
\frac{\bra{\psi^h_{g}}  [\rho_x^{n'm'} (k)^\dagger, [H_{eh},  \rho_{x}^{nm}(k) ] ]  \ket{\psi^h_{g}} }{\sqrt{\beta^{n'm'}_{k} \beta^{nm}_k}} &=&  -\sum_{q} \frac{V_q}{A}  e^{i \mathrm{Im}[q k^*]} G_{q^*}^{n'n}G_{-q^*}^{mm'} \frac {(1-\nu_{m'}) e ^{-|q|^2/2} - s_q^{m'm}}{\sqrt{(1-\nu_m)(1-\nu_{m'})}}, \\
\frac{\bra{\psi^h_{g}} [\rho_x^{n'm'} (k)^\dagger, [H_{hh},  \rho_{x}^{nm}(k) ] ]  \ket{\psi^h_{g}} }{\sqrt{\beta^{n'm'}_{k} \beta^{nm}_k}} &= &- \frac{\delta_{n'n}}{2}\sum_{q l} \frac{V_q}{A}  G_{q^*}^{ml}G_{-q^*}^{lm'}  \frac{s_q^{m'l}+s_{-q}^{lm'} +(\nu_l - \nu_m)e^{-|q|^2/2} }{\sqrt{(1-\nu_{m'})(1-\nu_{m})}} \label{eq24},
\end{eqnarray}
where we introduced: 
\begin{eqnarray}
s_q^{m'm}\equiv \frac{1}{N_\phi} \bra{\psi^h_{g}}  \rho_h^{mm'}(q) \rho_h^{m'm}(-q) \ket{\psi^h_{g}}.
\end{eqnarray}
We remark that the above derivation is completely general and that the only assumption made so far was that $\bra{\psi^h_{g}}  h^\dagger_{n'l'} h_{nl} \ket{\psi^h_{g}}=\nu_n \delta_{n'n}\delta_{l'l}$. We now make a further assumption that $\ket{\psi^h_{g}}$ is a Gaussian state, which allows us to use Wick's theorem to evaluate:
\begin{eqnarray}
s_q^{m'm}  = e^{-|q|^2/2}(1-\nu_{m'}) \nu_m. 
\end{eqnarray}
Since $s_q$ depends only on the magnitude of $q$, performing the angular integral over $\arg(q)$ in Eq.~(\ref{eq24}) will result in $\delta_{m'm}$. Using this and the expression for $s_q$, we can rewrite the above as: 
\begin{eqnarray}
\frac{\bra{\psi^h_{g}}  [\rho_x^{n'm'} (k)^\dagger, [H_0,  \rho_{x}^{nm}(k) ] ] \ket{\psi^h_{g}}}{\sqrt{\beta^{n'm'}_{k} \beta^{nm}_k}} &=&  \omega_c \left( n+m+1+\Delta_g/\omega_c\right) \delta_{n'n}\delta_{m'm}, \\
\frac{\bra{\psi^h_{g}} [\rho_x^{n'm'} (k)^\dagger, [H_{eh},  \rho_{x}^{nm}(k) ] ] \ket{\psi^h_{g}}}{\sqrt{\beta^{n'm'}_{k} \beta^{nm}_k}} &=&  -\sum_{q} \frac{V_q}{A}  e^{i \mathrm{Im}[q k^*]} G_{q^*}^{n'n}G_{-q^*}^{mm'} e ^{-|q|^2/2}  \sqrt{(1-\nu_m)(1-\nu_{m'})},\label{eq28}\\
\frac{\bra{\psi^h_{g}}  [\rho_x^{n'm'} (k)^\dagger, [H_{hh},  \rho_{x}^{nm}(k) ] ] \ket{\psi^h_{g}}}{\sqrt{\beta^{n'm'}_{k} \beta^{nm}_k}} &= &- \delta_{n'n}\delta_{m'm}\sum_{q l} \frac{V_q}{A}  G_{q^*}^{ml}G_{-q^*}^{lm}  e^{-|q|^2/2}\nu_l,\label{exchange} 
\end{eqnarray}
where the term in the second line represents the effects of the electron-hole interaction in the presence of the phase-space filling, while the term in the third line is due to hole-hole exchange (Fock) interaction. Notice that, in the case of a zero momentum exciton at $k=0$, the angular integral over $\arg(q)$ in Eq.~(\ref{eq28}) can be directly performed yielding a delta function $\delta_{n'-n-m'+m}$, which is a consequence of angular momentum conservation. Therefore, at $k=0$, the Hilbert space is divided into blocks labeled by the angular momentum $l=n-m$, out of which we are only interested in the optically bright $l=0$ subspace, whose ground state represents the experimentally-studied $1s$ state in the absence of the magnetic field~\cite{Mak_NatPhoton_2018}. Taking this into account and plugging the above formulas into Eq.~(\ref{eq20}), we can numerically compute the energy $\epsilon^\mathrm{X}_0$ of the exciton ground state at $B=16$~T as a function of the filling factor $\nu$ (we remark that in order to get convergence, we needed to include roughly $50$ LLs in the Hilbert space). The result of such a calculation including only the phase-space filling (i.e., $H_0+H_{eh}$), but ignoring other interactions (e.g., $H_{hh}$), is plotted with a black curve in Fig.~4(b) in the main text. The red curve in this figure presents the exciton energy dependence computed taking additionally into account the shifts arising due to hole-hole exchange interaction $H_{hh}$ and the exciton-hole scattering, which is described in the next section.

\subsection{Exciton-hole scattering}
In the previous section we have calculated the energy $\epsilon^X_0$ of the exciton ground state in a truncated Hilbert space, taking into account both electron-hole and hole-hole interactions. Now we consider the processes that take us outside of this single electron-hole pair Hilbert space, in order to include scattering of the exciton by itinerant holes. To this end, we assume that in the investigated low-hole-density regime ($a_B^*k_F\ll1$) the electron-hole Coulomb attraction remains strong enough, such that the exciton can be treated as a rigid particle. This allows us to focus only on the excitons occupying lowest-energy orbital state, which we calculated in the previous section. Furthermore, we treat the intravalley coupling between such excitons and the holes (arising due to virtual excitations to the higher-energy exciton states) as a contact interaction $U$ with a range determined roughly by the exciton Bohr radius. This interaction is also assumed to be repulsive, as we do not find any signatures of a spin-triplet, intravalley trion bound state in the experimental data.

\begin{figure}[t]
\includegraphics{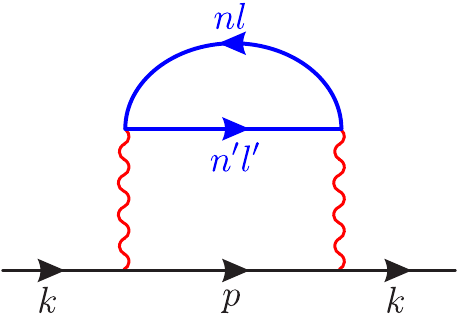}
\caption{Diagrammatic representation of the exciton self-energy in the second-Born approximation. The black/blue lines denote the exciton/hole propagators, while the red squiggly lines indicate the interaction between excitons and holes.\label{fig:theory_diagram}}
\end{figure}

Under the above assumptions, we may describe the interacting exciton-hole system with the following effective Hamiltonian:
\begin{eqnarray}
H_\mathrm{X-h}=\sum_k \epsilon^X_k X^\dagger_k X_k +\sum_{n l} \omega_c n  h_{nl}^\dagger h_{nl} + \frac{U}{A} \sum_{k}\sum_{|p|<\Omega} X^\dagger_{k+p} X_k \sum_{nln'l'} \bra{nl}e^{i\boldsymbol{p}\boldsymbol{r}_h}\ket{n'l'}h_{n'l'}^\dagger h_{nl},
\end{eqnarray}
which can be also expressed with the use of the hole density-like operator $\rho_h$ [defined by Eq.~(\ref{eq:rho_h})]:
\begin{eqnarray}
H_\mathrm{X-h}=\sum_k \epsilon^X_k X^\dagger_k X_k +\sum_{n} \omega_c n  \rho^{nn}_h(0) + \frac{U}{A} \sum_{k}\sum_{|p|<\Omega} X^\dagger_{k+p} X_k \sum_{nn'} G^{nn'}_{-p^*}  \rho^{n'n}_h(-p),
\end{eqnarray}
where $X_k$ ($X^\dagger_k$) represent the creation (anihilation) operator of the lowest-energy exciton of momentum $k$, while $\epsilon^X_k=\epsilon^X_0+|k|^2/{2 m^*_X}$ denotes the energy of this exciton state, with $m^*_X=m^*_e+m^*_h=2m^*$ standing for the exciton effective mass. The last term in the above Hamiltonian corresponds to the intravalley exciton-hole interaction, with $\Omega\approx1/a_B^*$ representing the ultra-violet cutoff that yields approximately $6.5/\ell_B$ at $B=16$~T. We calculate the self-energy energy acquired by the exciton due to this interaction term diagrammatically within a second-Born approximation (see Fig.~\ref{fig:theory_diagram}), which is justified for small $U$ (we neglect the first-Born contribution, as it does not contribute to the investigated filling-factor oscillations). Following Ref.~\cite{Efimkin_PRB_2018} (which once again makes the implicit assumption of a Gaussian hole state), it can be shown that the resulting self-energy $\Sigma(\omega,k=0)$ of the zero-momentum exciton is given by: 
\begin{equation}
\Sigma(\omega,k=0) =N_\phi \frac{U^2}{A^2} \sum_{nn'} \sum_{|p|<\Omega} \frac{\nu_n (1-\nu_{n'}) e^{-|p|^2/2} |G_p^{n'n}|^2}{\omega+ \omega_c n - \omega_c n' - \epsilon^X_{p} +i \gamma/2},  
\end{equation}
where $\gamma$ accounts for the finite exciton lifetime arising due to radiative broadening and disorder scattering (i.e., the FWHM linewidth of the exciton transition in the absence of exciton-hole interactions).

In the presence of the above self-energy, the retarded Green's function of the zero-momentum exciton is given by: 
\begin{equation}
G_x^R(\omega, k=0) = \frac{1}{\omega - \epsilon^X_0 - \Sigma(\omega,k=0) + i \gamma/2}.
\end{equation}
The shift of the $k=0$ exciton energy $\Delta\epsilon_\mathrm{int}$ due to the exciton-hole coupling can be readily obtained by determining the pole of the above propagator, i.e., by solving the following equation:
\begin{eqnarray}
\Delta\epsilon_\mathrm{int} - \mathrm{Re} [\Sigma(\Delta\epsilon_\mathrm{int}+\epsilon^X_0,k=0)] = 0.
\label{eq:delta_epsilon_Xh}
\end{eqnarray}
The total linewidth of the exciton transition in the presence of the interactions is in turn given by:
\begin{eqnarray}
\gamma_{\mathrm{total}} = \gamma - 2 \mathrm{Im} [\Sigma(\epsilon^X_0,k=0)] \equiv \gamma + \gamma_{\mathrm{int}},
\end{eqnarray}
where $\gamma_{\mathrm{int}}$ denotes the contribution to the linewidth arising from the exciton-hole coupling (notice that at this level of approximation the imaginary part of the self-energy should be evaluated on shell, i.e., for $\omega = \epsilon^X_0$). Since our model is intended to capture the LL-induced oscillatory features on a qualitative level only, in the following we will assume that $\gamma$ is small, and work in the limit $\gamma \to 0$ (a finite value for $\gamma$ will slightly smooth out the sharpness of the investigated oscillations). In this limit, we can explicitly evaluate the imaginary part of the self-energy using the relation $1/(x+i 0) \approx \mathrm{PV}(1/x) - i \pi \delta(x)$, where $\mathrm{PV}$ denotes the principal value: 
\begin{eqnarray}
\mathrm{Im}[\Sigma(\epsilon^X_0,k=0) ]&\approx& - \pi N_\phi \frac{U^2}{A^2} \sum_{nn'} \sum_{|p|<\Omega} \delta\left(n \omega_c - n' \omega_c - \frac{|p|^2}{2 m^*_X}\right)\nu_n (1-\nu_{n'})e^{-|p|^2/2} |G_p^{n'n}|^2\\&=&-\frac{m^*_X}{2} \frac{N_\phi}{A} U^2 \nu_{n_0} (1-\nu_{n_0}) = - \gamma_{\mathrm{int}}/2 \nonumber,
\end{eqnarray}
Here we took advantage of the fact that the factor $\nu_n (1-\nu_{n'})$ is non-zero only if $n' \geq n$, which in turn implies that the delta function can be satisfied only when $n'=n=n_0$, where $n_0$ represents the index of the last partially-filled LL. Using the above formula we can analytically compute the dependence of $\gamma_{\mathrm{int}}$ on the filling factor, which is plotted in Fig.~4(a) in the main text. For this calculation, the value of the exciton-hole interaction strength $U/\ell_B^2\approx7$~meV has been chosen manually in order for the oscillations amplitude $m^*_XN_\phi U^2/4A$ to remain consistent with that of $\sim1$~meV revealed by the experiment. Then, using the same value of $U$, we numerically compute the exciton-hole-interaction-induced exciton energy shift $\Delta\epsilon_\mathrm{int}$ from Eq.~(\ref{eq:delta_epsilon_Xh}), which we add to the exciton ground state energy $\epsilon^X_0$ of $H_0+H_{eh}+H_{hh}$ Hamiltonian determined in the previous section, and finally plot the resulting energy dependence as well as its derivative with the red curves in Figs~4(b,c) in the main text.